\documentclass[prd,preprint,superscriptaddress,nofootinbib,showpacs]{revtex4-2}

\usepackage[english]{babel}

\usepackage[utf8]{inputenc}
\usepackage{amsmath,amssymb}
\usepackage{bbold}
\usepackage{slashed}
\usepackage{subfigure}
\usepackage[colorlinks=true,
breaklinks=true,
urlcolor=blue,
citecolor=blue]{hyperref}
\usepackage[usenames,dvipsnames]{color}
\usepackage{appendix}   
\usepackage{braket,bm}
\usepackage{multirow}
\usepackage{enumitem}
\usepackage{color}
\usepackage{cancel}
\usepackage{dsfont}
\usepackage{orcidlink}
\usepackage{array}
\usepackage{indentfirst}
\usepackage{bbm}
\usepackage{epstopdf}
\usepackage{graphicx}
\usepackage{caption}

\DeclareCaptionLabelFormat{figprefix}{FIG.#2.} 
\allowdisplaybreaks          

\AtBeginDocument{}

\begin{document}

\title{Phase structure of 2+1-flavor QCD from an Einstein-dilaton-flavor holographic model}

\author{Jin-Yang Shen}
\email{Jinyang_Shen@hnu.edu.cn}
\affiliation{School of Physics and Electronics, Hunan University, Changsha 410082, China}

\affiliation{Hunan Provincial Key Laboratory of High-Energy Scale Physics and Applications, Hunan University, Changsha 410082, China}

\author{Xin-Yi Liu}
\email{liuxinyi23@mails.ucas.ac.cn}
\affiliation{School of Fundamental Physics and Mathematical Sciences, Hangzhou Institute for
 Advanced Study, UCAS, Hangzhou 310024, China}
 
\affiliation{Institute of Theoretical Physics, Chinese Academy of Sciences, Beijing 100190, China}

\affiliation{University of Chinese Academy of Sciences (UCAS), Beijing 100049, China}

\author{Jin-Rui Wu}
\email{wujinrui@hnu.edu.cn}
\affiliation{School of Physics and Electronics, Hunan University, Changsha 410082, China}

\author{Yue-Liang Wu}
\email{ylwu@itp.ac.cn}
\affiliation{School of Fundamental Physics and Mathematical Sciences, Hangzhou Institute for Advanced Study, UCAS, Hangzhou 310024, China}
\affiliation{Institute of Theoretical Physics, Chinese Academy of Sciences, Beijing 100190, China}
\affiliation{University of Chinese Academy of Sciences (UCAS), Beijing 100049, China}
\affiliation{International Center for Theoretical Physics Asia-Pacific (ICTP-AP), UCAS, Beijing 100190, China}

\author{Zhen Fang}
\email{zhenfang@hnu.edu.cn}
\affiliation{School of Physics and Electronics, Hunan University, Changsha 410082, China}

\affiliation{Hunan Provincial Key Laboratory of High-Energy Scale Physics and Applications, Hunan University, Changsha 410082, China}


\begin{abstract}
We construct a holographic QCD model based on the Einstein--dilaton--flavor framework with 2+1 flavors and investigate its phase structure using machine-learning techniques. At zero chemical potential, the model reproduces the equation of state and chiral transition in quantitative agreement with lattice QCD results. By varying the light and strange quark masses, we map out the quark-mass dependence of the transition order and obtain the corresponding phase diagram, which is consistent with phase structures extracted from lattice simulations and other nonperturbative approaches. In particular, the predicted first-order region is found to be small, in line with the most recent lattice QCD analyses. We also examine the critical behavior along the second-order boundaries and the tricritical region, finding that the critical exponents exhibit mean-field scaling characteristic of classical holographic constructions. Integrating machine learning with holographic QCD significantly enhances the efficiency of parameter optimization, providing a robust and practical strategy for improving the predictive power of holographic modeling of QCD thermodynamics.
\end{abstract}

\maketitle
\newpage

\section{Introduction}

Quantum chromodynamics (QCD), the fundamental theory of the strong interaction \cite{Gross:2022hyw,Gross:1973id,Politzer:1973fx}, predicts that at sufficiently high temperatures or baryon densities, quarks and gluons normally confined within hadrons become deconfined and form a novel state of strongly interacting matter known as the quark--gluon plasma (QGP) \cite{Cabibbo:1975ig,Yagi:2005quark,Collins:1974ky,Shuryak:1980tp}. Understanding the transition from hadronic matter to the QGP, as well as the global structure of the QCD phase diagram, remains a central goal in contemporary nuclear and particle physics \cite{Stephanov:2004wx,Busza:2018rrf}. Studies of QCD thermodynamics not only probe the nonperturbative behavior of the strong interaction but also shed light on the thermal evolution of the early universe, the dynamics of heavy-ion collisions, and the properties of dense matter in neutron stars \cite{Aoki:2006we,Fukushima:2025ujk}. Despite significant progress achieved through lattice QCD, effective field theories, and functional approaches, many aspects of the QCD phase transition, particularly the interplay between chiral symmetry restoration and deconfinement, remain incompletely understood and continue to motivate ongoing theoretical developments and refined modeling efforts.

\begin{figure}[htbp]
    \centering
\includegraphics[width=0.5\linewidth]{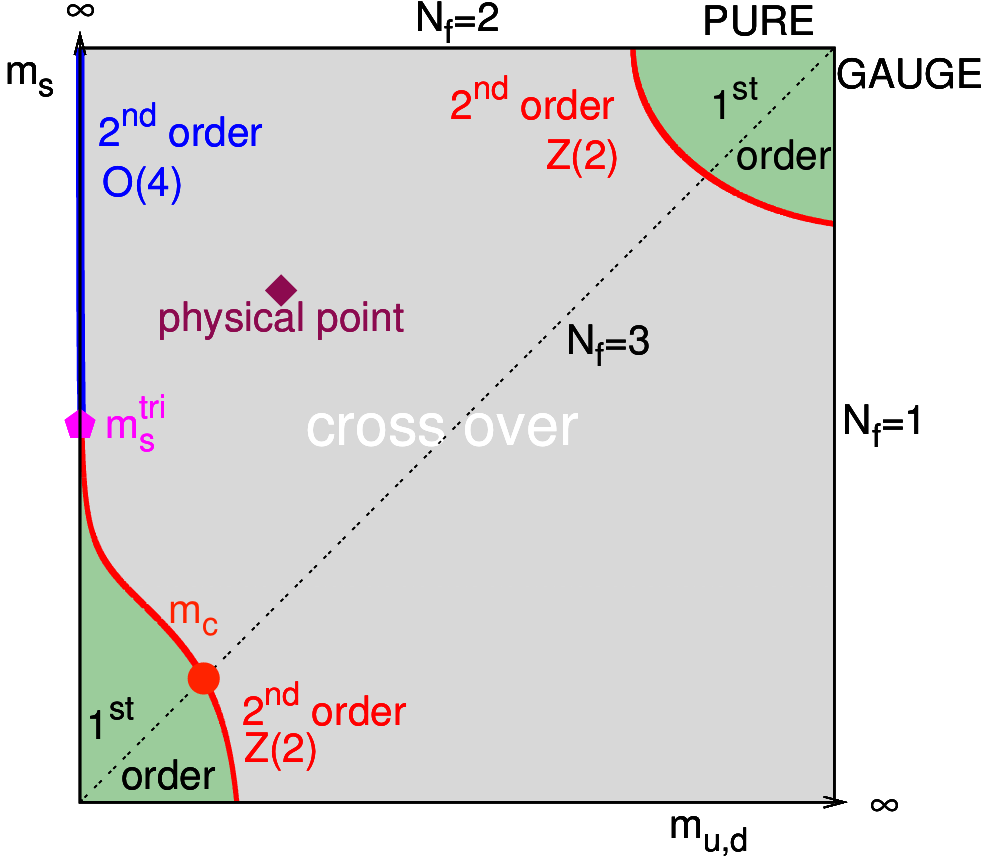}
    \caption{The quark-mass phase diagram (Columbia plot) illustrating how the QCD transition order at $\mu =0$ depends on $m_{u,d}$ and $m_s$ \cite{Ding:2015ona}.}
    \label{fig:1}
\end{figure}

Among the theoretical approaches to the QCD phase transition, lattice QCD and effective model studies have played essential and complementary roles. At vanishing baryon chemical potential $(\mu = 0)$, extensive lattice simulations have mapped out how the nature of the thermal transition depends on the light and strange quark masses~\cite{Fischer:2014ata,Gao:2020qsj,Fukushima:2008wg,HotQCD:2014kol,Borsanyi:2013bia}. This dependence is summarized by the quark-mass phase diagram, known as the Columbia plot~\cite{Brown:1990ev,Laermann:2003cv}, shown in Fig.~\ref{fig:1}. At the physical point, the QCD transition is firmly established as a smooth crossover. In the chiral limit ($m_{u,d}\to 0$ with fixed and large $m_s$), the transition is expected to become second order in the O(4) universality class, while in the heavy-quark limit it turns first order due to the approximate restoration of $Z(3)$ center symmetry. The Columbia plot thus encapsulates the symmetry structure that controls the order of the QCD thermal transition and serves as a guide for theoretical explorations into regions that remain challenging for first-principle calculations.

Motivated by the AdS/CFT correspondence \cite{Witten:1998qj,Gubser:1998bc,Maldacena:1997re}, holographic QCD provides a complementary nonperturbative tool for studying strongly coupled gauge dynamics. Over the past two decades, a variety of bottom-up holographic constructions, such as hard-wall, soft-wall, and V-QCD models etc, have demonstrated notable success in describing hadron spectra, confinement, and chiral symmetry breaking properties \cite{DaRold:2005mxj,Erlich:2005qh,Karch:2006pv,deTeramond:2005su,Babington:2003vm,He:2007juu,Gherghetta:2009ac,Cherman:2008eh,Kelley:2010mu,Jarvinen:2011qe,Jarvinen:2015ofa,Chelabi:2015cwn,Chelabi:2015gpc,Li:2016smq,Sui:2009xe,Sui:2010ay,Cui:2013xva,Li:2012ay,Li:2013oda,Fang:2016nfj,Rougemont:2023gfz,Giannuzzi:2025fsv,Mager:2025pvz,Si:2025pry,Zheng:2024rzl,Zheng:2025qcg,Jokela:2024xgz}. These models have since been refined to incorporate dynamical backgrounds and thermodynamic observables, enabling studies of the QCD phase transition and its sensitive dependence on quark masses \cite{Li:2022erd,Liu:2023pbt,Jarvinen:2025mgj}. Nonetheless, many existing frameworks do not dynamically couple the gluonic and flavor sectors, limiting their ability to simultaneously describe deconfinement and chiral restoration. This naturally raises the question whether one can construct a unified holographic framework that self-consistently encodes both transitions and reproduces the full structure of the Columbia plot.

Recently, rapid developments in machine learning and artificial intelligence have introduced powerful optimization and data-driven techniques into high-energy and nuclear physics \cite{Zhou:2023pti}. In the context of holographic QCD, machine-learning methods have been applied to parameter optimization, reconstruction of bulk geometries, and systematic matching to lattice data \cite{Akutagawa:2020yeo,Hashimoto:2018bnb,Hashimoto:2018ftp,Yan:2020wcd,Chen:2024mmd,Hashimoto:2021ihd,Song:2020agw,Li:2022zjc,Cai:2024eqa,Chen:2024ckb}. These approaches significantly accelerate the exploration of high-dimensional parameter spaces and improve the predictive stability of holographic models. Building on these developments, we employ neural-network–assisted optimization to determine the parameters of an Einstein--dilaton--flavor (EDF) model in an efficient and self-consistent manner. This strategy is particularly advantageous because the EDF setup contains a set of coupled parameters that jointly influence the thermodynamic response, chiral dynamics, and infrared structure of the background geometry. By training the optimization network against both the QCD equation of state (EoS) and the temperature-dependent chiral condensates, we ensure that the gluonic and flavor sectors are simultaneously constrained within a unified holographic framework. With the optimized model in hand, we compute the quark-mass phase diagram, investigate the structure of its first-order, second-order, and crossover regions, and analyze in detail the location of the tricritical point as well as the behavior along the flavor-symmetric line, extending and refining the results previously outlined in Ref. \cite{Shen:2025zkj}.

A further aspect relevant to the construction of the Columbia plot concerns the characterization of critical behavior along second-order boundaries and in the vicinity of the tricritical region \cite{Pisarski:1983ms,Stephanov:2006zvm,Rajagopal:1992qz,Cuteri:2021ikv}. The location and nature of second-order boundaries are quite important for the critical endpoint on the QCD phase diagram \cite{Fodor:2001pe,Fodor:2004nz,deForcrand:2006pv}, which is one of the primary objectives of  heavy-ion collision experiments. The critical exponents extracted from our model exhibit Landau-type mean-field scaling, which is expected for holographic models formulated at the classical gravity level, where critical fluctuations are intrinsically suppressed and the dynamics near the transition are governed predominantly by smooth background fields \cite{DeWolfe:2011ts,DeWolfe:2010he,Chen:2018vty,Chen:2018msc,Fu:2024nmw,Zhang:2025vsm}. Although mean-field behavior does not reproduce the nonperturbative universality classes predicted for QCD, such as the O(4) class anticipated in the two-flavor chiral limit or the Z(2) class near the endpoint of the first-order region, the resulting scaling patterns still provide valuable qualitative information. In particular, they allow us to clearly locate the boundaries separating crossover, second-order, and first-order regions, and to identify the emergence of a tricritical point in the quark-mass plane. In this sense, even mean-field critical scaling serves as a meaningful diagnostic tool for mapping the topology of the phase diagram within the holographic framework.

This paper is organized as follows. In Sec.~\ref{sec:2}, we present the components of the EDF model and derive the associated equations of motion (EOMs) that govern the coupled dynamics of the metric, dilaton, and flavor fields. Sec.~\ref{sec:3} introduces our machine-learning optimization strategy and details the computation of the chiral condensates and equation of state. Sec.~\ref{sec:4} presents the resulting quark-mass phase diagram, highlights the properties of the tricritical point and the flavor-symmetric case, and discusses the critical exponents along the different segments of the critical line. In Sec.~\ref{sec:5}, we summarize our main results and discuss possible directions for future investigation.

\section{The Einstein-dilaton-flavor system}\label{sec:2}

\subsection{The bulk action}

In previous studies \cite{Liu:2023pbt,Liu:2024efy}, the QCD phase transition has been investigated within a two-flavor holographic framework based on the Einstein–Maxwell–dilaton system coupled with an improved soft-wall model. In the present work, we extend this framework to the 2+1-flavor case at zero chemical potential, aiming to explore both chiral and deconfinement transitions at finite temperatures. We begin by adopting the following metric ansatz in the Einstein frame:
\begin{equation}\label{metric1}
    \begin{split}
        ds^2 = \frac{L^2 e^{2 A_E(z)}}{z^2} \left(-f(z)dt^2 + \frac{dz^2}{f(z)} + dx^i dx^i\right),
    \end{split}
\end{equation}
where \( i \) denotes the spatial index, with \( i = 1, 2, 3 \). \( A_E(z) \) represents the warp factor, and \( L \) is the curvature radius of the asymptotically AdS\(_5\) spacetime, which can be conveniently set to unity. The holographic radial coordinate is denoted by \( z \), where \( z \to 0 \) corresponds to the asymptotic AdS boundary, and \( z = z_h \) marks the event horizon, at which the black hole condition \( f(z_h) = 0 \) must be satisfied.

To construct the holographic model, we extend the two-flavor action employed in previous studies to incorporate the \( 2 + 1 \) flavor case~\cite{Fang:2019lmd, Fang:2018axm, Fang:2018vkp}, leading to an Einstein–dilaton–flavor (EDF) system described by the following action:
\begin{equation}\label{Stotal}
    \begin{split}
        S=&\frac{1}{2 \kappa_N^2}\int d^5x\sqrt{-g}\left[R -\frac{4}{3}(\partial \phi)^2 -V_E(\phi) \right.\\
         &\left. - \beta e^{\phi}\left((\partial\chi_u)^2+\frac{1}{2}(\partial\chi_s)^2 +V(\chi_u,\chi_s,\phi)\right) \right],
    \end{split}
\end{equation}
where \( \kappa_N = \sqrt{8 \pi G_5} \), and \( \beta \) represents the coupling between the matter fields and the gravitational background. Without loss of generality, we set \( \beta = 1 \) by appropriately rescaling the vacuum expectation values (VEVs) of the scalar fields \( \chi_u \) and \( \chi_s \). The potential \( V(\chi_u, \chi_s, \phi) \), which couples the light and strange quark sectors to the dilaton field, is expressed as:
\begin{align}
    V(\chi_u,\chi_s,\phi)= e^{\frac{4\phi}{3}} V_\chi(\chi_u,\chi_s,\phi),
\end{align}
where
\begin{equation}\label{Vchichiusphi}
    V_\chi(\chi_u,\chi_{s},\phi) =-\frac{1}{2} \left(3+\Phi(\phi)\right)(2\chi_u^2+\chi_s^2) + \frac{\gamma}{2\sqrt{2}}\chi_u^2\chi_s +\frac{\lambda}{4} (2\chi_u^4+\chi_s^4) .
\end{equation}
For the coupling function \( \Phi(\phi) \), we adopt the specific form \( \Phi(\phi) = d_1 \phi + d_2 \phi^2 \)~\cite{Liu:2023pbt}. 

By applying a scale transformation \( \phi_c = \sqrt{8/3}\,\phi \), the kinetic term of the dilaton field \( \phi \) can be normalized to its canonical form. Following Ref.~\cite{Gubser:2008yx}, the dilaton potential is then expressed as
\begin{equation}
    V_c(\phi_c) = \frac{1}{L^2} \left( -12 \cosh \gamma_1 \phi_c + b_2 \phi_c^2 + b_4 \phi_c^4 \right),
\end{equation} 
We adopt this form in our model by identifying \( V_E(\phi) = V_c(\phi_c) \). This ensures that the bulk geometry maintains an asymptotic AdS structure with \( \Lambda = -6 \) near the boundary:
\begin{equation} 
 V_c(\phi_c \to 0) \simeq -\frac{12}{L^2} + \frac{b_2 - 6 \gamma_1^2}{L^2} \phi_c^2 + \mathcal{O}(\phi_c^4). 
\end{equation} 
The mass-dimension relation then gives \begin{equation} 
b_2 = 6 \gamma_1^2 + \frac{\Delta (\Delta - 4)}{2}, 
\end{equation} 
where \( \Delta \) denotes the scaling dimension of the operator dual to the dilaton field. For simplicity, we set \( \Delta = 3 \). The detailed construction of the action~(\ref{Stotal}) is provided in Appendix \ref{app1}.

\subsection{The EOM}

Starting from the action (\ref{Stotal}), we derive the Einstein equation and the EOMs for the dilaton $\phi$ and the scalar VEVs $\chi_u$ and $\chi_s$ as follows:
\begin{align}
    & R_{MN} -\frac{1}{2} g_{MN}R +\frac{4}{3} \left(\frac{1}{2} g_{MN}\partial_J\phi\partial^J\phi -\partial_M\phi\partial_N\phi\right) + \frac{1}{2} g_{MN}V_E(\phi)   \notag\\
    & +\frac{\beta}{2} g_{MN}e^{\phi}\, V(\chi_u, \chi_s, \phi) +\beta e^{\phi} \left(\frac{1}{2} g_{MN} \partial_J\chi_u \partial^J\chi_u -\partial_M\chi_u \partial_N\chi_u\right)   \notag\\
    & +\frac{\beta e^{\phi}}{2} \left(\frac{1}{2} g_{MN} \partial_J\chi_s \partial^J\chi_s -\partial_M\chi_s \partial_N\chi_s\right) =0 ,   \\
    &\frac{8}{3}\nabla_M \nabla^M \phi -\partial_\phi V_E(\phi) -\beta \partial_\phi (e^{\phi} V(\chi_u,\chi_s, \phi))   \notag\\ 
 &-\beta e^{\phi} g^{MN} \left(\partial _M \chi _u\partial_N \chi_u  +\frac{1}{2} \partial_M \chi_s\partial_N \chi_s\right) =0,   \\
&\nabla_M (e^{\phi} \nabla^M\chi_u) - e^{\phi} \partial_{\chi_u} V(\chi_u, \chi_s, \phi) =0,
\\
&\nabla_M (e^{\phi} \nabla^M \chi_s) - e^{\phi}\partial_{\chi_s} V(\chi_u, \chi_s, \phi) =0.
\end{align}
Substituting the metric ansatz \eqref{metric1} into the above system and assuming that all fields depend only on the radial coordinate \( z \), the system reduces to a set of five independent ordinary differential equations:
\begin{equation}\label{eom1}
    \begin{split}
        &f''-\frac{3f'}{z}+3A_E'f'=0,
    \end{split}
\end{equation}
\begin{equation}
    \begin{split}
        A_E''-A_E'^2+\frac{2A_E'}{z}+\frac{4}{9}\phi'^2+\frac{1}{3}\beta e^{\phi}\chi_u'^2+\frac{1}{6}\beta e^{\phi}\chi_s'^2=0,
    \end{split}
\end{equation}
\begin{align}
        &\phi''+\left(\frac{f'}{f}+3A_E'-\frac{3}{z}\right) \phi' -\frac{3e^{2A_E} \partial_\phi V_E(\phi)}{8 z^2 f}  \notag\\
        &-\frac{3\beta}{8}e^\phi \chi_u'^2-\frac{3\beta}{16}e^\phi \chi_s'^2 -\frac{3\beta e^{2A_E} \partial _\phi (e^\phi V(\chi_u,\chi_s, \phi))}{8 z^2 f} =0,
\end{align}
\begin{equation}
    \begin{split}
        \chi_u''+\left(\frac{f'}{f}+3A_E'-\frac{3}{z}+\phi'\right)\chi_u'-\frac{e^{2A_E}\partial_{\chi_u} V(\chi_u,\chi_s,\phi)}{2 z^2 f} =0,
    \end{split}
\end{equation}
\begin{equation}\label{eom5}
    \begin{split}
        \chi_s''+\left(\frac{f'}{f}+3A_E'-\frac{3}{z}+\phi'\right)\chi_s'-\frac{ e^{2A_E}\partial_{\chi_s} V(\chi_u,\chi_s,\phi)}{z^2 f} =0.
    \end{split}
\end{equation}

\subsection{The boundary condition}

To solve the EDF system, appropriate boundary conditions must be imposed. For the function \( f(z) \), we adopt the following boundary conditions:
\begin{align}
    f(0)=1,\qquad  f(z_h)=0.
\end{align}

The remaining boundary conditions can be derived from the ultraviolet (UV) asymptotic behavior of the EOMs (\ref{eom1})--(\ref{eom5}). Near the AdS boundary, the fields admit the following asymptotic expansions:
\begin{align}
    f(z) &= 1 - f_4 z^4 + ...\,,   \label{UV_f}\\
    A_E(z) &= \frac{1}{108}(-8 p_1^2 -3(2m_{u}^2+m_s^2)\beta \zeta^2)z^2   \notag\\
        &\quad +a_3 z^3 +a_4 z^4 +a_{4l} z^4 \ln z +...\,,   
     \\
    \phi(z) &= p_1 z + \frac{3}{16} \beta\zeta^2 (2 m_u^2+m_s^2)(6+d_1 ) z^2 +p_3 z^3   \notag\\
        &\quad +p_{3l} z^3 \ln z +p_4 z^4 +p_{4l} z^4 \ln z +...\,,
\\
    \chi_u (z) &= m_u \zeta z -\frac{1}{4} m_u \zeta (\sqrt{2} m_s \gamma \zeta -4 p_1 (5 + d_1 )) z^2 +\frac{\sigma_u}{\zeta}z^3   \notag\\ 
    &\quad +c_{3l} z^3 \ln z +c_{4}z^4 +c_{4l}z^4 \ln z +...\,,
\\
    \chi_s (z) &= m_s \zeta z - \frac{1}{4}\zeta (\sqrt{2}m_u^2 \gamma \zeta - 4 m_s p_1 (5 + d_1 )) z^2 +\frac{\sigma_s}{\zeta}z^3   \notag\\ 
    &\quad +\tilde{c}_{3l} z^3 \ln z +\tilde{c}_{4}z^4 +\tilde{c}_{4l}z^4 \ln z +...\,,   \label{UV_chi_s}
\end{align}
where \( m_u \) and \( m_s \) denote the light and strange quark masses, respectively, while \( \sigma_u \) and \( \sigma_s \) correspond to the chiral condensates, as discussed below. The parameter \( \zeta \) is a normalization constant \cite{Cherman:2008eh}. As pointed out in Ref. \cite{Fang:2016nfj}, the precise value of \( \zeta \) remains uncertain; therefore, in this work we treat \( \zeta \) as a free parameter and set \( \zeta = 1 \) for convenience. Different choices of \( \zeta \) can equivalently be compensated by adjusting the coupling parameter \( \beta \). The explicit expressions for the subleading UV coefficients are provided in Appendix \ref{app2}.

From the above expansions, we obtain the following boundary conditions at the AdS boundary:
\begin{align}\label{BC1}
    \phi'(0) =p_1,\qquad  \chi_u'(0) =m_u \zeta,\qquad  \chi_s'(0) =m_s \zeta.
\end{align}
Compared with the two-flavor case, the \( 2 + 1 \)-flavor system involves an additional scalar field \( \chi_s(z) \). The coefficients \( \sigma_u \) and \( \sigma_s \) are extracted from the UV asymptotics of \( \chi_u(z) \) and \( \chi_s(z) \), respectively. To numerically solve the five coupled differential equations~(\ref{eom1})--(\ref{eom5}), we employ the spectral method.

\section{Chiral transition and thermodynamic properties} \label{sec:3} 

We investigate the QCD phase transition in the \( 2 + 1 \)-flavor case by numerically solving the EDF system to obtain the EoS and the chiral condensates. This allows for a detailed analysis of the thermodynamic properties and phase structure of the system at zero chemical potential. The temperature \( T \) and entropy density \( s \) are given by
\begin{equation}
    T = \frac{|f'(z_h)|}{4\pi}, \qquad 
    s = \frac{2\pi e^{3 A_E(z_h)}}{\kappa_N^2 z_h^3}.
\end{equation}
The pressure \( p \) and energy density \( \varepsilon \) follow from the standard thermodynamic relations:
\begin{equation}
    dp= sdT,\qquad \varepsilon =-p+sT.
\end{equation}

According to the holographic dictionary, the chiral condensates for the light and strange quarks are expressed as
\begin{equation}\label{consensateR}
    \langle\Bar{\psi} \psi\rangle_{u}^T =\frac{\delta S_{r}}{\delta m_{u}} = \frac{\beta}{\kappa_N^2} \sigma_{u} +b_1, \qquad \langle\Bar{\psi} \psi\rangle_{s}^T = \frac{\delta S_{r}}{\delta m_{s}} = \frac{\beta}{\kappa_N^2} \sigma_{s} + b_2,
\end{equation}
where \( b_1 \) and \( b_2 \) denote the functional contributions to the renormalized condensates. The full expressions for 
\( \langle \bar{\psi} \psi \rangle_u^T \) and \( \langle \bar{\psi} \psi \rangle_s^T \) are obtained via holographic renormalization using the on-shell action \( S_r \equiv (S + S_\partial)_{\text{on-shell}} \), where \( S_\partial \) represents the boundary terms. The detailed derivations are provided in Appendix \ref{app3}.

To compare the normalized chiral condensates with the results of lattice QCD for \( 2 + 1 \) flavors, we adopt the fitting formulas from the HotQCD lattice collaboration \cite{Gubler:2018ctz}:
\begin{align}
\frac{\langle \bar{u} u \rangle_T}{\langle 0 | \bar{u} u | 0 \rangle} = 1 - \frac{\hat{d} - \Delta_u^R(T)}{\hat{d} - \Delta_u^R(\infty)}, \qquad \frac{\langle \bar{s} s \rangle_T}{\langle 0 | \bar{s} s | 0 \rangle} = 1 + \frac{\hat{d} - \Delta_s^R(T)}{2 m_s r_1^4 \langle 0 | \bar{s} s | 0 \rangle}, \label{eq:hotQCD_s}
\end{align}
where $\Delta_q^R = \hat{d} + 2m_s r_1^4 \left[\langle\bar{\psi} \psi\rangle_q^T - \langle\bar{\psi} \psi\rangle_q^0\right]$ with parameters $\hat{d}=0.0232244$ and $r_1=0.3106$ fm. The vacuum strange quark condensate is taken as $\langle0| \bar{s}s|0\rangle=-(0.307~\text{GeV})^3$. The zero-temperature condensate $\langle\Bar{\psi}\psi\rangle_{q}^0$ can be obtained either by fitting it as a free parameter to lattice data or by evaluating the finite-temperature condensate $\langle\Bar{\psi}\psi\rangle_{q}^T$ at sufficiently low temperatures \cite{Fu:2024nmw,Cai:2022omk}.

In addition, we refer to the Wuppertal–Budapest (WB) lattice results and compute the subtracted and renormalized chiral condensates as \cite{Borsanyi:2010bp}
\begin{align}\label{eq:WB}
\Delta_{l,s} = \frac{\langle \bar{\psi} \psi \rangle_u^T - \frac{m_u}{m_s} \langle \bar{\psi} \psi \rangle_s^T}{\langle \bar{\psi} \psi \rangle_u^0 - \frac{m_u}{m_s} \langle \bar{\psi} \psi \rangle_s^0}, \qquad \langle\bar{\psi}\psi\rangle_R=-\frac{m_u}{X^4}[\langle\bar{\psi}\psi\rangle_{u}^T-\langle\bar{\psi}\psi\rangle_{u}^0],
\end{align}
where $X$ is a reference scale introduced for normalization and conveniently set to be the pion mass. For the purpose of fitting the lattice data, we choose $X = 0.102~\text{GeV}$.

\subsection{A slice of machine learning and neural network}

To obtain self-consistent numerical solutions within the EDF framework, the model parameters must be carefully calibrated using first-principles lattice QCD data for the 2+1-flavor system. In particular, to simultaneously capture both the chiral and deconfinement transitions, the parameter optimization must satisfy two key constraints: reproducing the EoS and the temperature-dependent chiral condensates. This ensures that the gravitational dual faithfully encodes the nonperturbative thermodynamics and critical behavior across both the confinement and chiral symmetry breaking regimes.

There are seven parameters, $\gamma, \gamma_1, b_4, \lambda, d_1, d_2$, and the Newton constant $G_5$ in the action~(\ref{Stotal}), as well as three parameters, $p_1, m_u, m_s$, appearing in the boundary conditions~(\ref{BC1}). Due to the high dimensionality of this parameter space and the necessity of simultaneous fitting to the above observables, traditional trial-and-error methods are computationally inefficient and unscalable. Therefore, we employ neural-network-based machine learning optimization algorithms to determine the parameter set.

\begin{figure}[htbp]
    \centering
\includegraphics[width=0.6\linewidth]{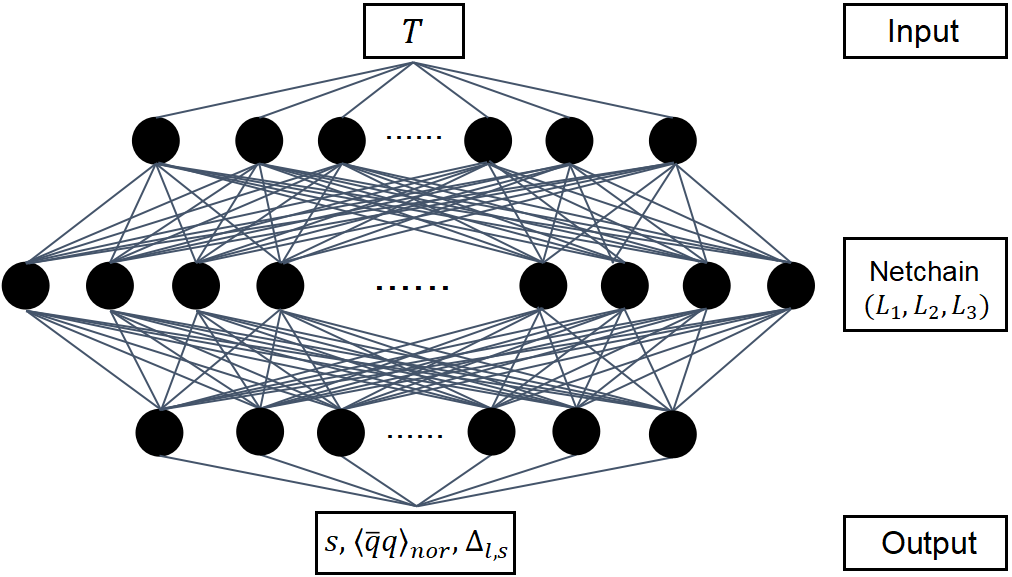}
    \caption{Schematic diagram of the neural network architecture. The network consists of a single-input layer, three hidden layers with 128, 256, and 128 neurons, respectively, and a single-output layer.}
    \label{fig:2}
\end{figure}

We combine machine learning with the spectral method implemented in \textit{Mathematica} to optimize the parameters. Specifically, as illustrated in Fig.~\ref{fig:2}, we use the temperature $T$ in the lattice data (for $s$–$T$, $\langle\bar{q} q\rangle_{\text{nor}}$–$T$, and $\Delta_{l,s}$–$T$) as the input, while the entropy density $s$, normalized chiral condensate $\langle\bar{q} q\rangle_{\text{nor}}$, and subtracted chiral condensate $\Delta_{l,s}$ serve as the outputs. The deep neural network consists of one input layer, three hidden layers $L_1, L_2, L_3$ (with 128, 256, and 128 neurons per layer, respectively, each using ReLU activation), and one output layer. The mean squared error is adopted as the loss function, and the Adam optimization algorithm is employed for training. After 10,000 training epochs, the network yields smooth interpolating functions for the observables, enabling highly accurate prediction and extrapolation beyond the available lattice data.

Subsequently, we employ the gradient descent algorithm combined with the Adam optimizer to automatically refine the model parameters. The mean squared error between the neural network’s fitted outputs and the holographic QCD model’s predicted observables is computed and adopted as the loss function. Since the procedure requires solving differential equations numerically, a numerical gradient must be constructed. Let $Loss(x)$ denote the loss function and $x$ a model parameter; then, under a small perturbation $\epsilon$, the numerical gradient of the loss function with respect to $x$ is defined as
\begin{align}
    Gradient \equiv \frac{Loss(x+\epsilon)-Loss(x-\epsilon)}{2 \epsilon}.
\end{align}

Through this integrated optimization framework, we achieve an automated and high-precision calibration of model parameters, yielding a self-consistent fit to lattice QCD data for both thermodynamic and chiral observables.

\subsection{Machine learning-optimized parameters and fitting results}

By performing a global fit to lattice QCD data, the quark masses that best reproduce QCD thermodynamics are determined to be $m_{u,d}^{\text{phy}} = 3.5~\text{MeV}$ and $m_s^{\text{phy}} = 139~\text{MeV}$. The corresponding values of the remaining model parameters are listed in Table \ref{table:1}. To properly account for the coupling between the light and strange quark sectors, two additional parameters, $\gamma$ and $d_2$, are introduced beyond those present in the two-flavor case \cite{Liu:2023pbt}.
\begin{table}[htbp]
  \centering
  \begin{tabular}{c c c c c c c c}
    \hline\hline
     $\gamma_1$ & $b_4$ & $G_5$ & $\gamma$ & $\lambda$ & $d_1$ & $d_2$ & $p_1$ \\
    \hline
     0.737 & 0.12 & 0.48 & 1.7 & 2.4 & -0.86 & -0.115 & 0.365 \\
    \hline\hline
  \end{tabular}
  \caption{Parameter values of the 2+1-flavor EDF model determined via machine learning optimization. The parameter $p_1$ is given in units of GeV.}
  \label{table:1}
\end{table}

\begin{figure}[htbp]
    \centering
\includegraphics[width=0.6\linewidth]{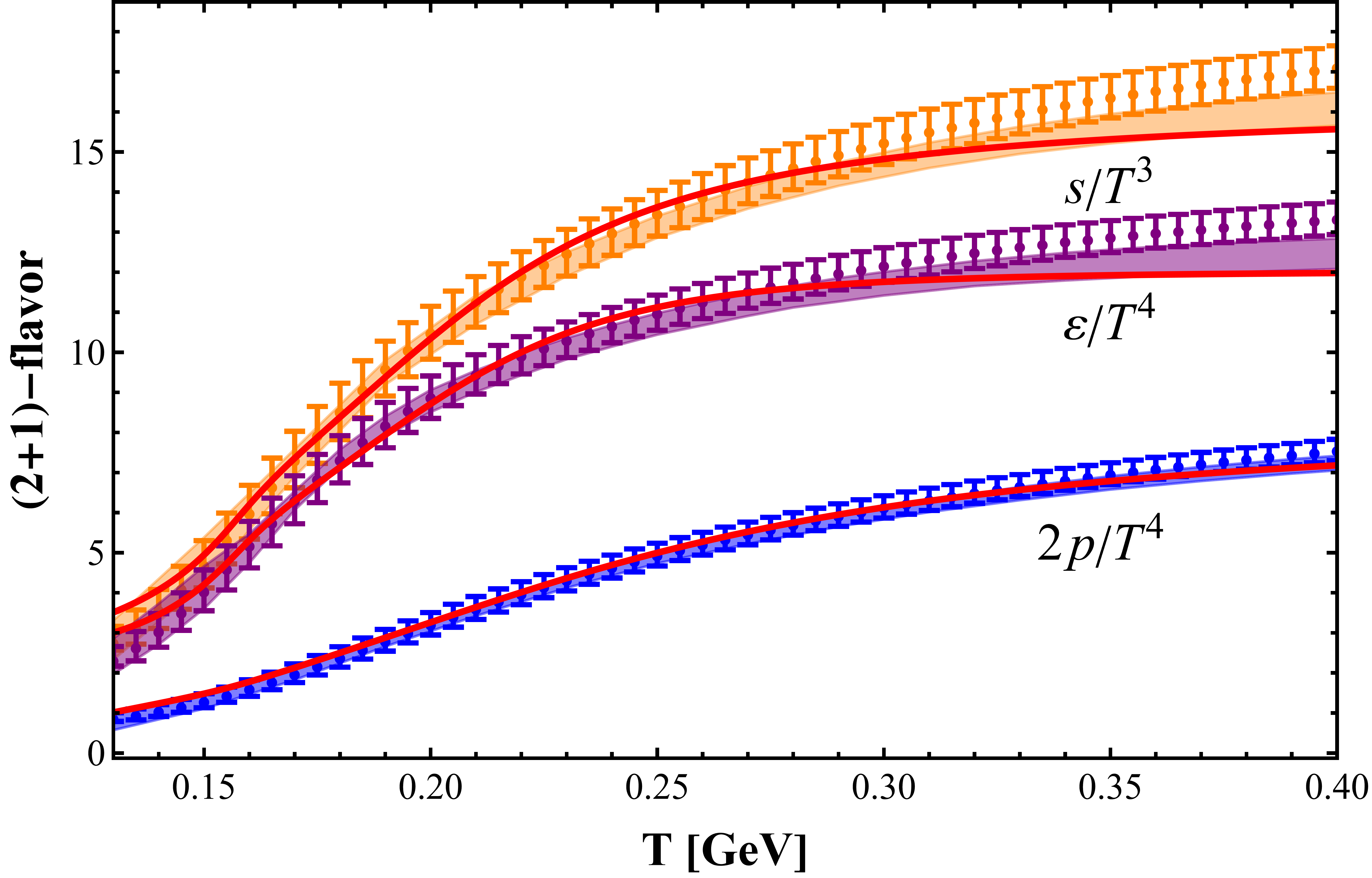}
\caption{Temperature dependence of the energy density $\varepsilon$, entropy density $s$, and pressure $p$ in the 2+1-flavor case. The results from the machine learning–optimized holographic model are shown as solid lines, while the points with error bars and shaded bands correspond to the HotQCD~\cite{HotQCD:2014kol} and W-B~\cite{Borsanyi:2013bia} lattice simulations, respectively.}
    \label{fig:3}
\end{figure}
\begin{figure}[htbp]
    \centering
\includegraphics[width=0.62\linewidth]{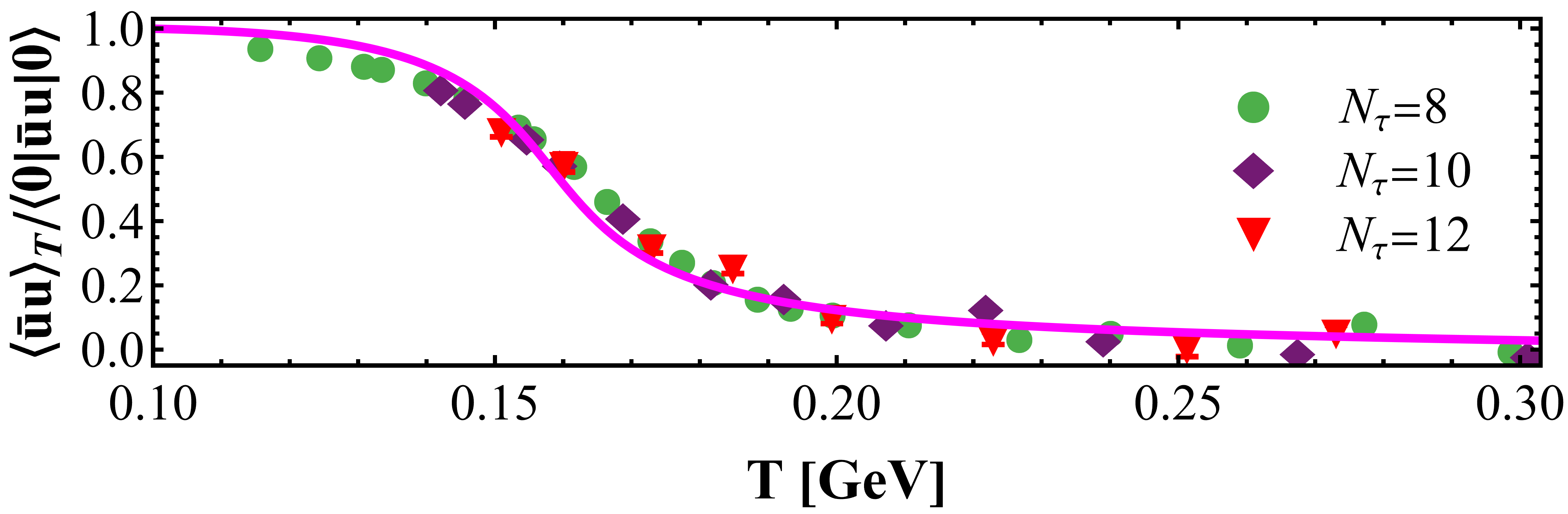}
\includegraphics[width=0.62\linewidth]{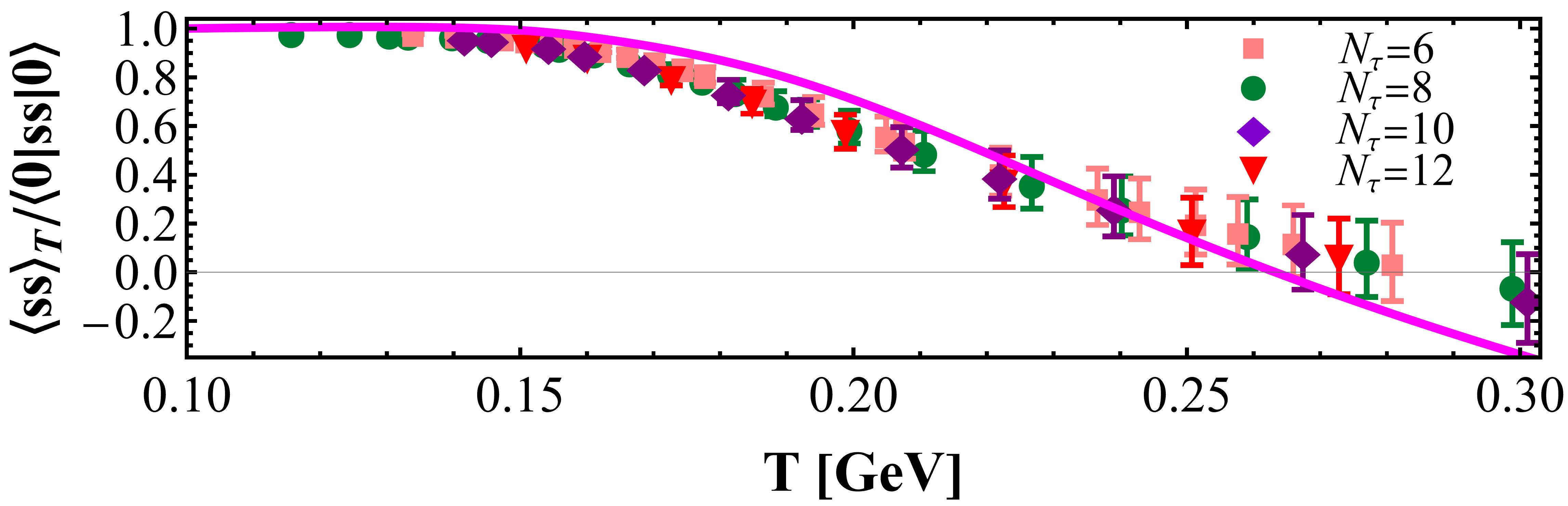}
\caption{Comparison of the chiral transition behavior for light (top) and strange (bottom) quarks with the HotQCD results~\cite{HotQCD:2014kol,Gubler:2018ctz}.}
    \label{fig:4}
\end{figure}
\begin{figure}[htbp]
    \centering
\includegraphics[width=0.62\linewidth]{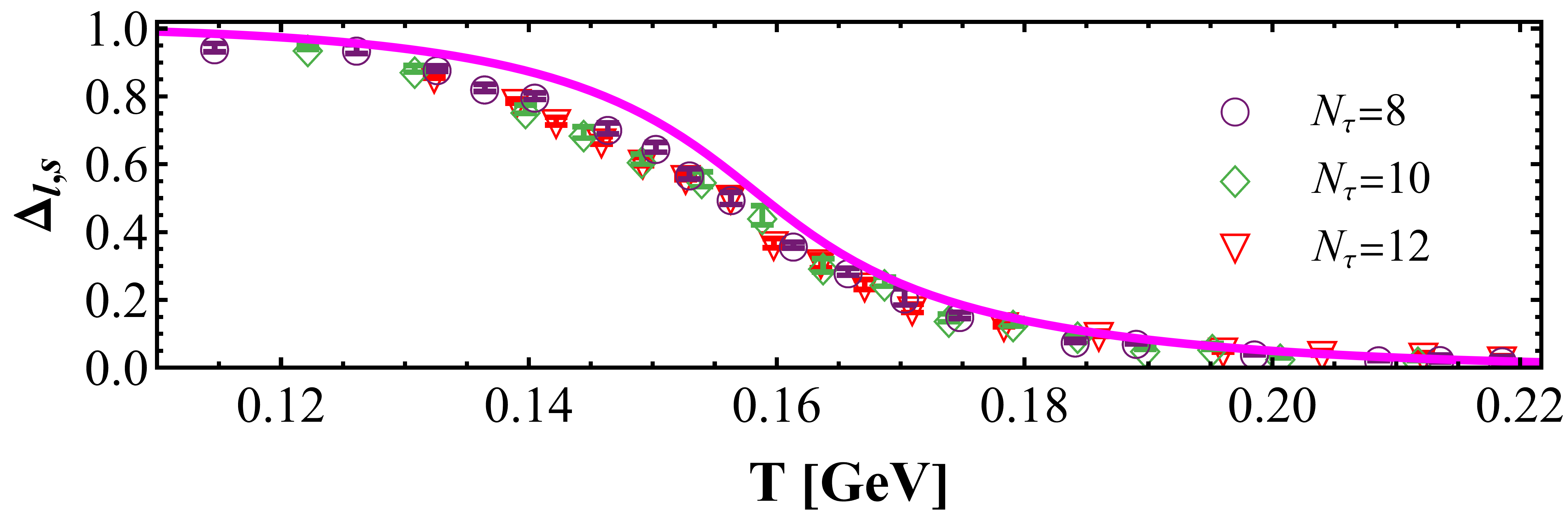}
\includegraphics[width=0.62\linewidth]{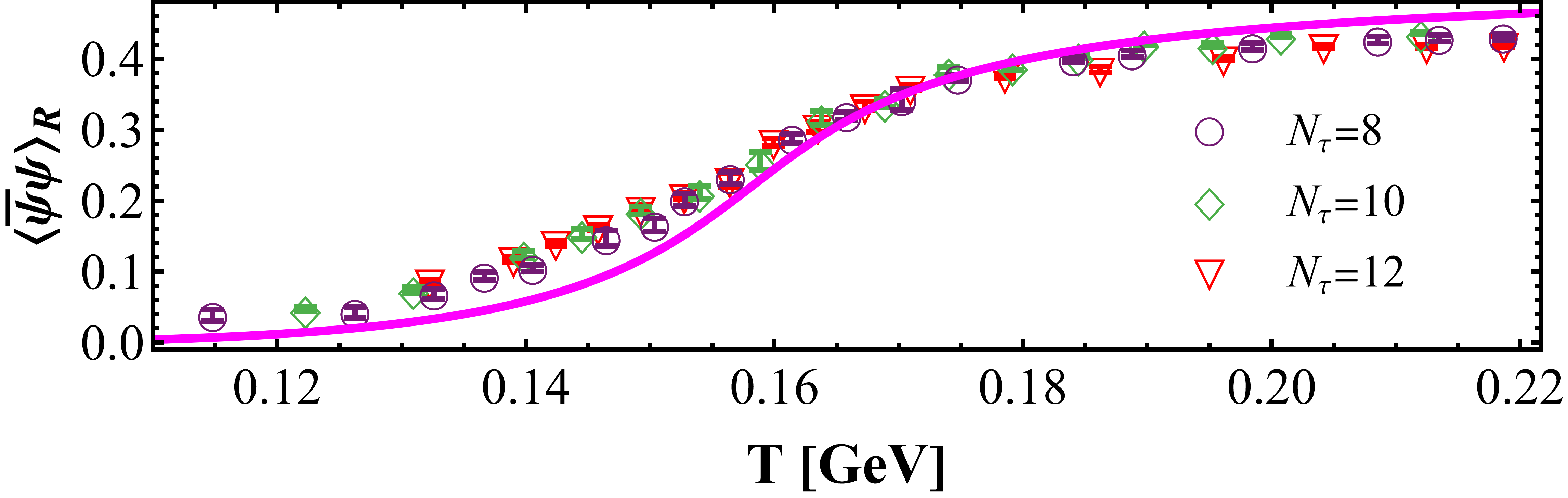}
\caption{The subtracted chiral condensate (top) and the renormalized chiral condensate (bottom) compared with the W-B lattice results~\cite{Borsanyi:2010bp}.}
    \label{fig:WBsigma}
\end{figure}

The thermodynamic quantities at $\mu = 0$ obtained from our model and lattice QCD are presented in Fig.~\ref{fig:3}, including the energy density $\varepsilon$, entropy density $s$, and pressure $p$. The chiral transition behaviors for the light and strange quarks are shown in Fig.~\ref{fig:4}, the subtracted chiral condensate and the renormalized chiral condensate are shown in Fig.~\ref{fig:WBsigma}. The corresponding lattice data are taken from Refs.~\cite{Borsanyi:2013bia,HotQCD:2014kol,Gubler:2018ctz,Borsanyi:2010bp}. It is evident that both the EoS and the normalized quark condensates computed from our 2+1-flavor improved soft-wall model exhibit good agreement with the lattice results. It is worth noting that, since the model parameters are calibrated directly against lattice QCD data, the parameter-fitting procedure itself fixes the physical normalization, so that the holographic thermodynamic quantities can be compared with lattice observables without introducing additional scaling factors.

In contrast to the method employed in Ref.~\cite{Cai:2022omk}, we couple the flavor sector to the gravitational background and use a unified parameter set to simultaneously reproduce both the EoS and the chiral condensates. Although this strategy considerably increases the fitting complexity, it enhances the physical consistency and predictive reliability of the model, thereby offering clearer insights into the correlations between thermodynamic observables and chiral dynamics.

\subsection{Pseudo-critical temperatures of the QCD thermal transition}

We have obtained the EoS and chiral transition behavior that are consistent with lattice QCD results. To further validate the holographic QCD model and gain deeper insight into the thermal transition, we analyze the temperature dependence of the order parameters for both deconfinement and chiral restoration, and determine the associated pseudo-critical temperatures.

\begin{figure}[htbp]
    \centering
    \includegraphics[width=0.6\linewidth]{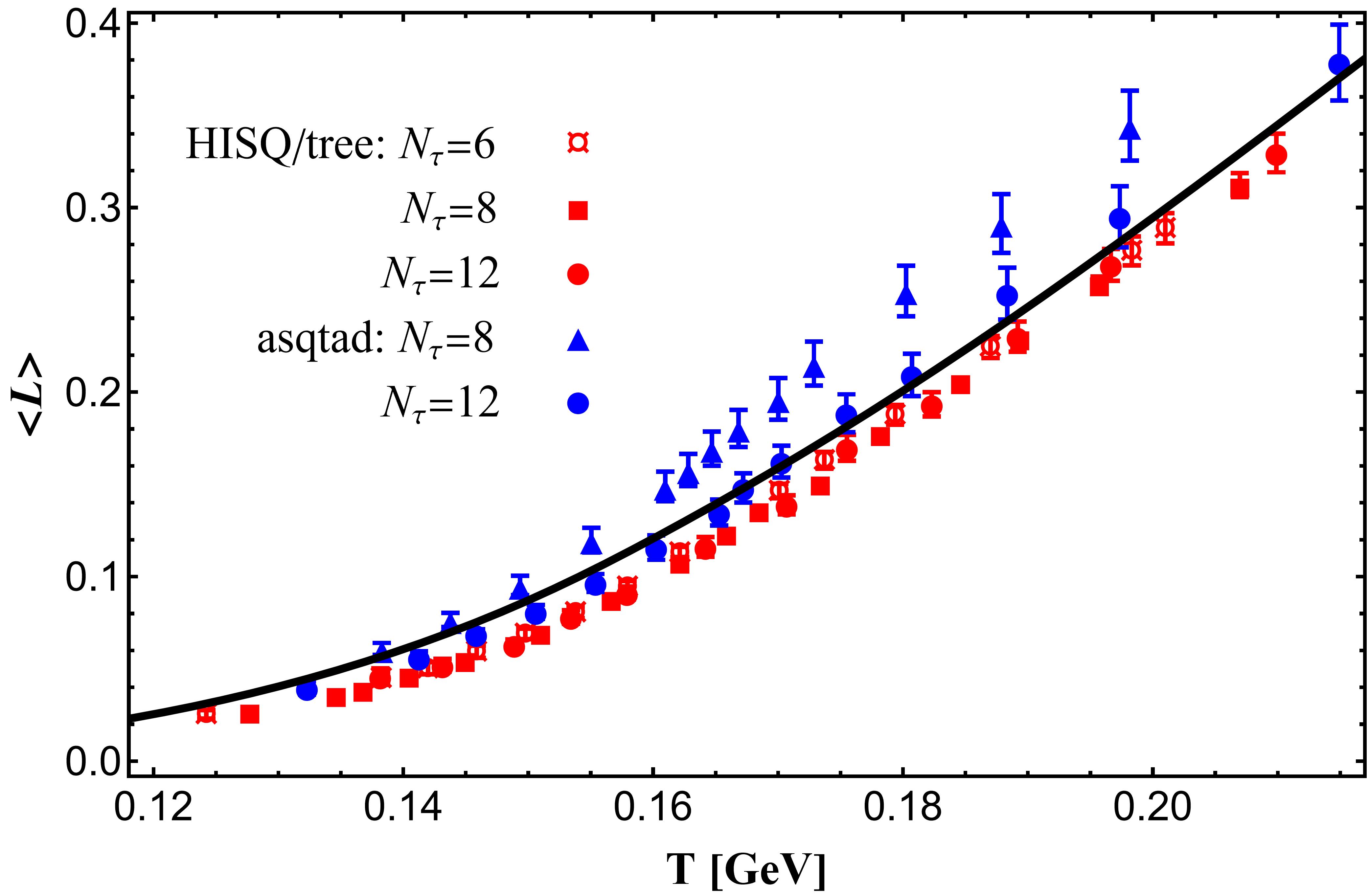}
    \caption{The expectation value of the Polyakov loop, $\langle L(T) \rangle$. The black solid line represents the model results, while the points with error bars denote the 2+1-flavor lattice data taken from Ref. \cite{Bazavov:2011nk}.}
    \label{polyloop1}
\end{figure}

The deconfining transition temperature is typically extracted from the behavior of the Polyakov loop, which serves as a key indicator of deconfinement. In the holographic framework, the expectation value of the Polyakov loop is computed as \cite{Andreev:2009zk,Fang:2015ytf} 
\begin{equation}
\langle L \rangle = e^{c_p - S_0},
\end{equation} 
where $c_p$ is a normalization constant and $S_0$ is given by
\begin{equation} 
S_0 = \frac{g_p}{\pi T} \int_{0}^{z_h} dz \left( \frac{e^{2 A_S}}{z^2} - \frac{1}{z^2} \right),
\end{equation} 
with the string-frame warp factor defined as $A_S = A_E + \tfrac{2}{3}\phi$ and $g_p = \tfrac{L^2}{2 \alpha_p}$, where $\alpha_p$ represents the string tension. To avoid divergence at $z = 0$, the lower integration limit is set to $10^{-4}$. We treat $c_p$ and $g_p$ as free parameters and determine their optimal values by fitting to lattice QCD data, obtaining $c_p = 1.3$ and $g_p = 0.31$. Figure~\ref{polyloop1} presents a comparison between our results and lattice data from Ref.~\cite{Bazavov:2011nk}.

The pseudo-critical temperatures corresponding to chiral restoration and deconfinement are then determined. The chiral transition temperature, \(T_\chi\), is defined as the temperature at which the derivative of the light quark chiral condensate with respect to temperature $\left|d\langle\bar{\psi}\psi\rangle_u/dT\right|$ attains its maximum, yielding \(T_\chi \simeq 158.2~\text{MeV}\). This value agrees closely with the lattice QCD result \(T_\chi^{\text{lattice}} = (156.5 \pm 1.5)~\text{MeV}\) \cite{HotQCD:2018pds}. The deconfining transition temperature, \(T_d\), is extracted from the inflection point of the renormalized Polyakov loop, giving \(T_d \simeq 228.9~\text{MeV}\). For comparison, lattice QCD studies with dynamical quarks typically report values in the range \(170\text{--}200~\text{MeV}\) \cite{Aoki:2009sc,Bazavov:2009zn,Bazavov:2016uvm}. It should be emphasized that, while the Polyakov loop serves as a true order parameter only in the pure-gauge limit (\(T_d^{\text{gauge}} \simeq 285~\text{MeV}\) \cite{Borsanyi:2022xml,Giusti:2025fxu}), it remains a valuable probe of deconfinement in the presence of dynamical quarks \cite{Bazavov:2011nk}. 

Additionally, the temperature at which the squared speed of sound, \(c_s^2\), reaches its minimum is found to be \(T_c \simeq 155.5~\text{MeV}\), consistent with the lattice QCD estimate \(T_c = 154 \pm 9~\text{MeV}\) \cite{HotQCD:2014kol}. The slight differences among these characteristic temperatures reflect the crossover nature of the QCD transition at vanishing chemical potential, where no true thermodynamic singularity exists and different observables exhibit peak structures at distinct temperatures.

The hierarchy \(T_\chi < T_d\) indicates that the chiral restoration and deconfinement transitions occur at different temperatures in our model. This separation may suggest the presence of an intermediate regime in which quarks become active degrees of freedom while gluonic dynamics remain partially confined. It has been proposed that such an intermediate phase may correspond to a state with approximate chiral symmetry restoration through quark liberation, while gluons remain bound in glueball-like excitations until higher temperatures are reached \cite{Cohen:2023hbq,Fujimoto:2025sxx}. This decoupling between chiral and deconfinement transitions highlights the potential existence of hybrid quark–gluon phases and motivates further exploration within the holographic QCD framework.

\section{Mass dependence of the QCD phase transition}\label{sec:4}

\begin{figure}[htbp]
    \centering   \includegraphics[width=0.45\linewidth]{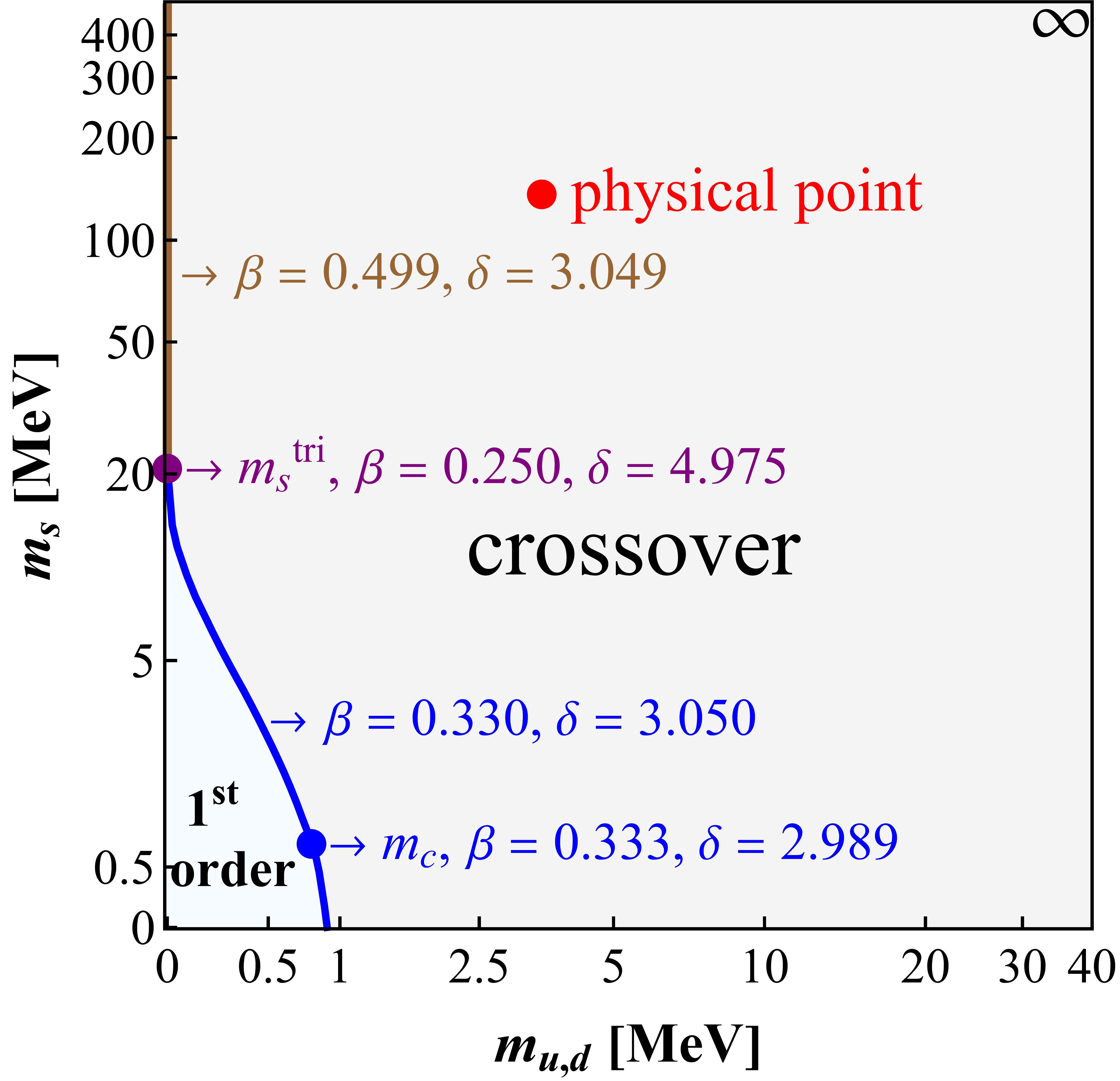}
    \caption{Columbia phase diagram obtained from the 2+1-flavor EDF system. The blue and brown solid curves represent the chiral critical line that separates the first-order phase transition region from the crossover region. The symbols $\beta$ and $\delta$ denote the critical exponents at various locations along the chiral critical line. The quark masses at the physical point are $m_{u,d}^{\text{phy}} = 3.5$ MeV and $m_s^{\text{phy}} = 139$ MeV.}
    \label{fig:5}
\end{figure}

With the model parameters determined through machine learning optimization against lattice QCD data, we proceed to investigate the quark-mass dependence of the chiral and deconfinement phase structures in the 2+1-flavor system. Our primary goal is to reproduce the Columbia phase diagram, as illustrated in Fig.~\ref{fig:1}. The results of our analysis are presented in Fig.~\ref{fig:5}, excluding the pure gauge sector located in the upper-right corner. In this diagram, the horizontal axis represents the light quark mass $m_{u,d}$, and the vertical axis denotes the strange quark mass $m_s$, both plotted on logarithmic scales to capture the behavior across a broad range of quark masses. Three representative benchmark points are indicated:  
\begin{itemize}
    \item \textbf{Physical point} (red point): $m_{u,d}^{\text{phy}} = 3.5$~MeV and $m_s^{\text{phy}} = 139$~MeV.
    \item \textbf{tricritical point} $m_s^{\text{tri}}$ (purple point): $m_{u,d}=0$~MeV and $m_s=21.2$~MeV.
    \item \textbf{Flavor-symmetric point} $m_c$ (blue point): $m_{u,d}=m_s=0.785$~MeV.
\end{itemize}

Our model predictions are qualitatively consistent with the lattice QCD expectations reported in Ref.~\cite{Ding:2015ona}, as shown in Fig. \ref{fig:5}. In particular, we confirm that the strange quark mass at the tricritical point satisfies $m_s^{\text{tri}} < m_s^{\text{phy}}$. It is noteworthy that in our earlier work~\cite{Fang:2018vkp}, the same problem was studied using an improved 2+1-flavor soft-wall model, but only the chiral transition was considered under a fixed gravitational background. That approach prevented a direct comparison with lattice QCD results and led to the opposite conclusion, $m_s^{\text{tri}} > m_s^{\text{phy}}$. The present study demonstrates that coupling the flavor sector dynamically to the gravitational background is essential for reproducing the realistic QCD phase structure.

Furthermore, we extract the critical exponents along different segments of the chiral critical line: $\beta = 0.499$ and $\delta = 3.049$ on the brown segment, $\beta = 0.330$ and $\delta = 3.050$ on the blue segment, $\beta = 0.250$ and $\delta = 4.975$ at the tricritical point, and $\beta = 0.333$ and $\delta = 2.989$ at the flavor-symmetric point.


\subsection{The tricritical point}

In the chiral limit of two massless light quarks ($m_{u,d} \to 0$), renormalization group analyses predict a second-order QCD phase transition that belongs to the O(4) universality class when the QCD axial anomaly is strong, or a first-order transition when the anomaly is substantially weakened \cite{Pisarski:1983ms}. For a QCD system with three massless quark flavors, the chiral transition is generally expected to be of first order. When $m_s = m_s^{\text{phy}}$ and $m_{u,d} \to 0$, lattice QCD simulations indicate a second-order transition in the O(4) universality class \cite{Ejiri:2009ac}, although the strength of the chiral anomaly near the critical temperature remains uncertain. Along the $m_{u,d} = 0$ boundary, a second-order phase transition occurs at a specific strange quark mass $m^{\text{tri}}_s$, corresponding to the tricritical point. The precise value of this mass is still unknown, and even the possibility that $m^{\text{tri}}_s \to \infty$ cannot be ruled out.

\begin{figure}[htbp]
    \centering  \includegraphics[width=0.55\linewidth]{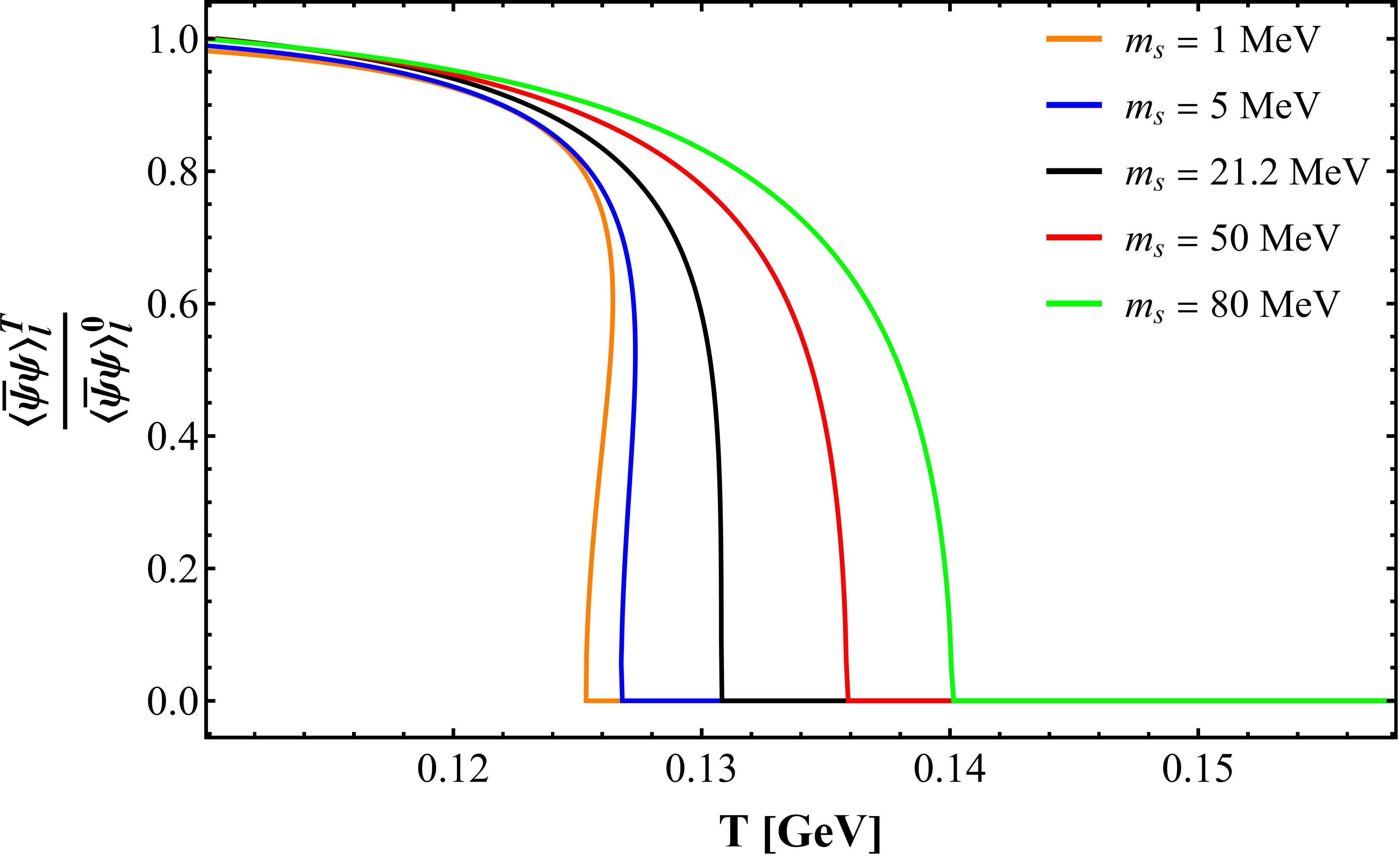}
    \caption{The temperature dependence of the normalized light-quark chiral condensates at $m_s = 1$, 5, 21.2, 50, and 80 MeV in the $m_{u,d} = 0$ case.}
    \label{fig:6}
\end{figure}
\begin{figure}[htbp]
    \centering
\includegraphics[width=0.55\linewidth]{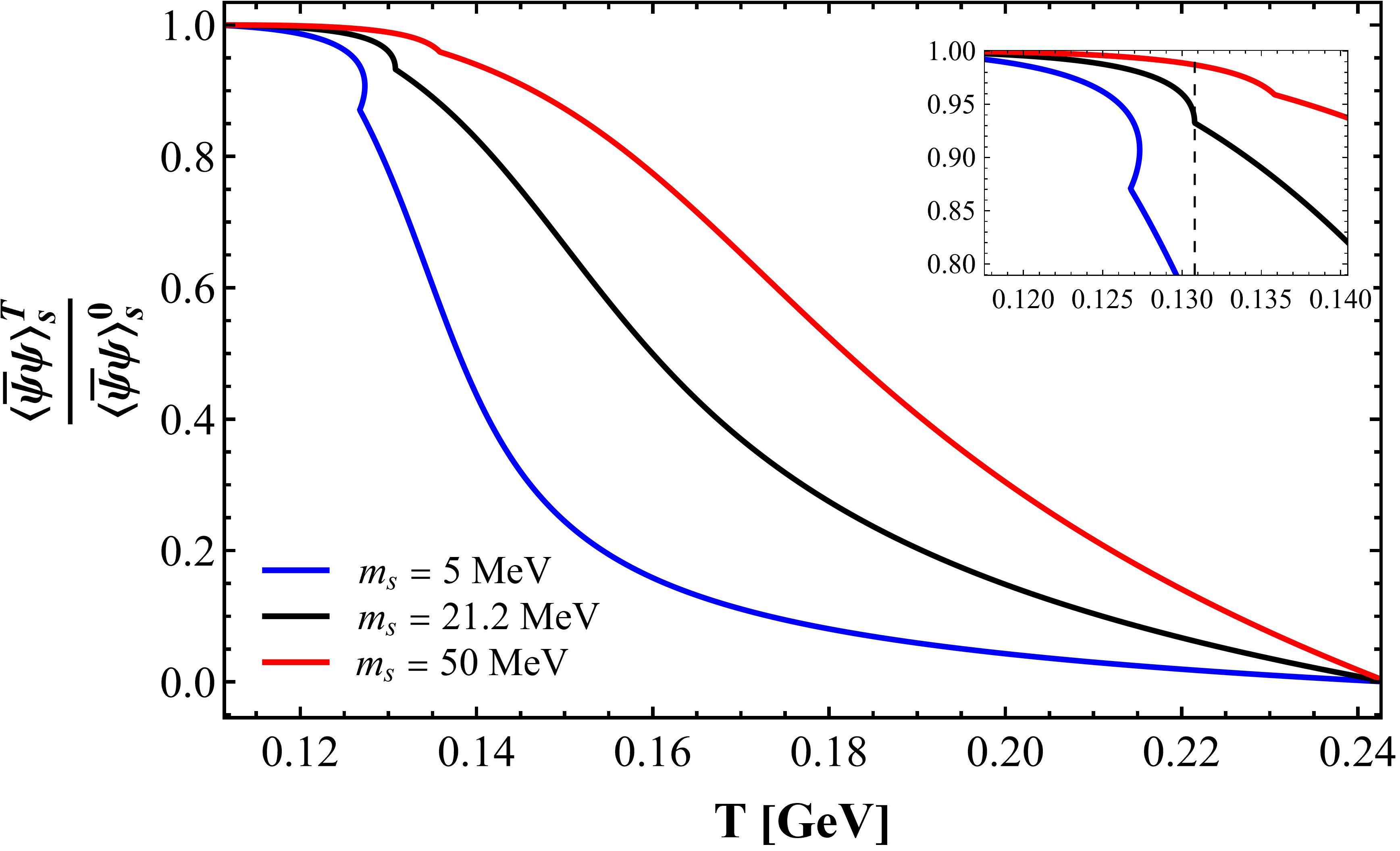}
\caption{The temperature dependence of the normalized chiral condensate of the strange quark at $m_s = 5$, $21.2$, and $50$ MeV in the $m_{u,d}=0$ case.}
    \label{fig:7}
\end{figure}

To determine the tricritical strange quark mass $m_s^{\text{tri}}$ in our model, we fix $m_{u,d} = 0$ and vary the strange quark mass $m_s$ across a broad range to compute the normalized chiral condensates and thermodynamic observables. By analyzing the resulting data and corresponding plots, we identify the value of $m_s$ at which the phase transition changes from first-order to second-order, marking the tricritical point. Our calculation yields $m_s^{\text{tri}} = 21.2$~MeV, which is smaller than the physical strange quark mass $m_s^{\text{phy}}$. 

\begin{figure}[htbp]
    \centering
\includegraphics[width=0.49\linewidth]{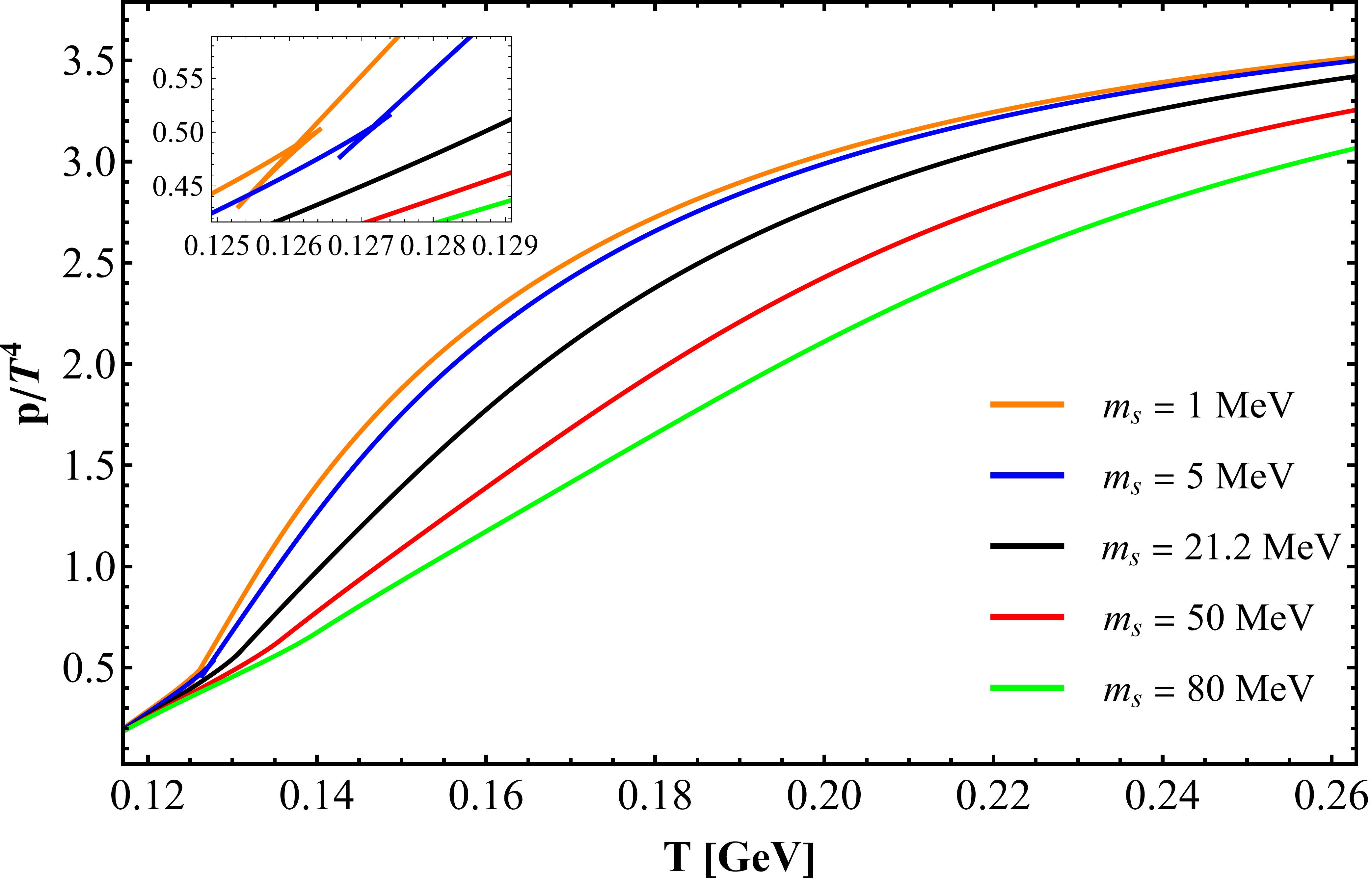}
\includegraphics[width=0.49\linewidth]{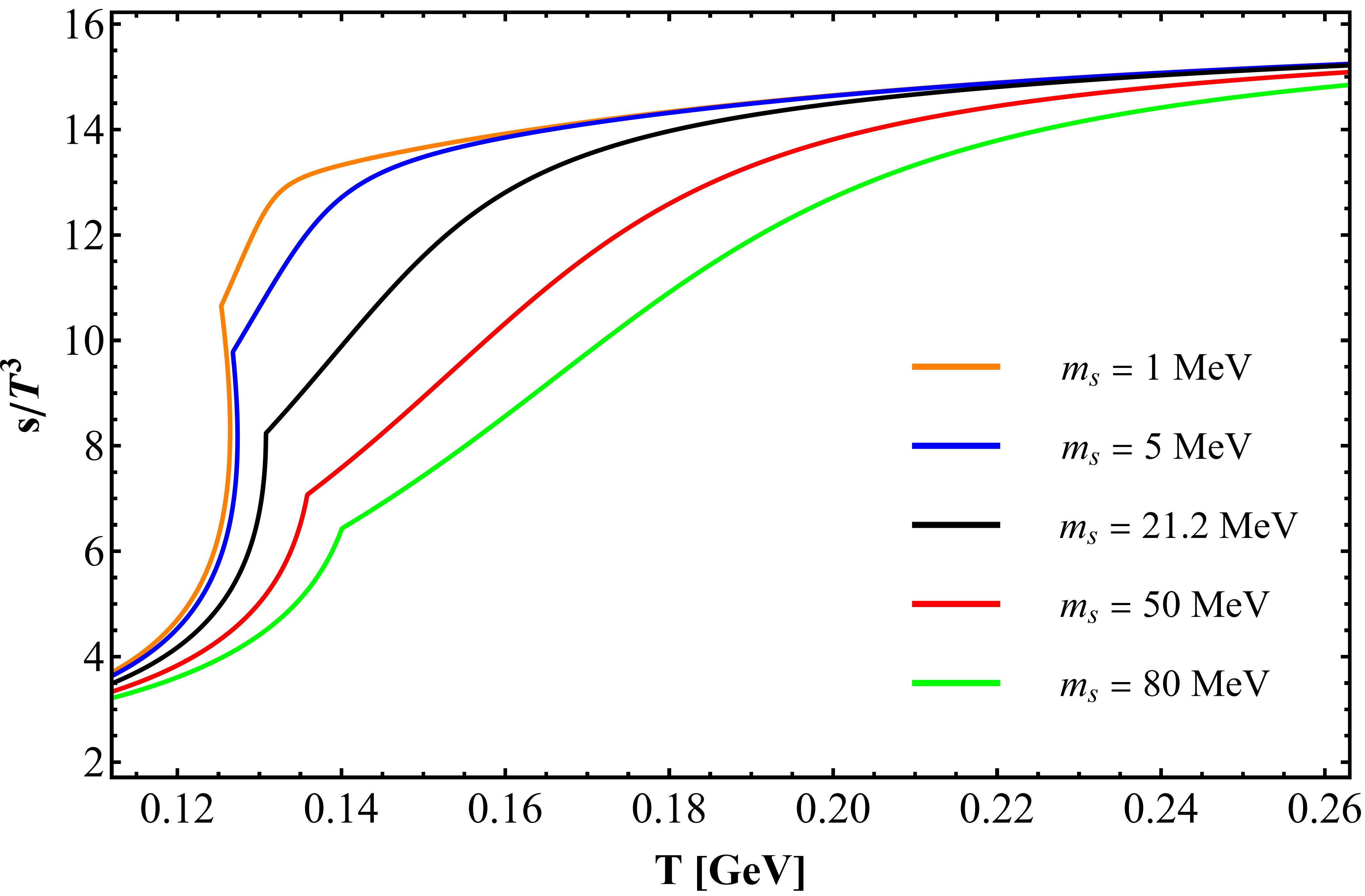}
\caption{The temperature dependence of the pressure $p$ and entropy density $s$ at $m_s = 1$, $5$, $21.2$, $50$, and $80$ MeV in the $m_{u,d}=0$ case.}
    \label{fig:8}
\end{figure}

The numerical results for the normalized chiral condensates of the $u$ and $s$ quarks, along with the pressure $p$ and entropy density $s$ as functions of temperature $T$, for several representative values of $m_s$, are presented in Figs.~\ref{fig:6}, \ref{fig:7}, and \ref{fig:8}. As shown in Fig.~\ref{fig:6}, for small strange quark masses, the light-quark chiral condensate (depicted by the yellow and blue lines) exhibits a pronounced inflection point at lower temperatures, corresponding to a clear dovetail structure in the left panel of Fig.~\ref{fig:8}. 

Comparing the temperature dependence of the chiral condensates for the light and strange quarks reveals a marked contrast: the light-quark condensate decreases sharply near the critical temperature, whereas the strange-quark condensate exhibits a more gradual decline. The temperatures at which chiral symmetry is restored for the two quark flavors differ significantly, reflecting the impact of the quark mass hierarchy. As the strange-quark mass decreases, its condensate also diminishes more rapidly. Furthermore, as illustrated in Fig.~\ref{fig:8}, the system pressure $p$ and entropy density $s$ for different $m_s$ values converge at higher temperatures. This convergence signifies chiral symmetry restoration and the transition from the confined phase to the QGP phase as the temperature increases.

\subsection{The flavor-symmetric case}

In the flavor-symmetric case ($m_u = m_d = m_s$), we find that for small quark masses, two distinct sets of VEV solutions for $\chi_u$ and $\chi_s$ emerge within a narrow temperature window near the critical region. These correspond to two solution branches:  
\begin{itemize}
    \item \textbf{Case a:} $\chi_u \neq \chi_s$ near the phase transition temperature, leading to distinct chiral condensates for the light and strange quarks even under flavor symmetry.  
    \item \textbf{Case b:} $\chi_u$ and $\chi_s$ remain identical across the entire temperature range.
\end{itemize}

\begin{figure}[htbp]
    \centering
\includegraphics[width=0.49\linewidth]{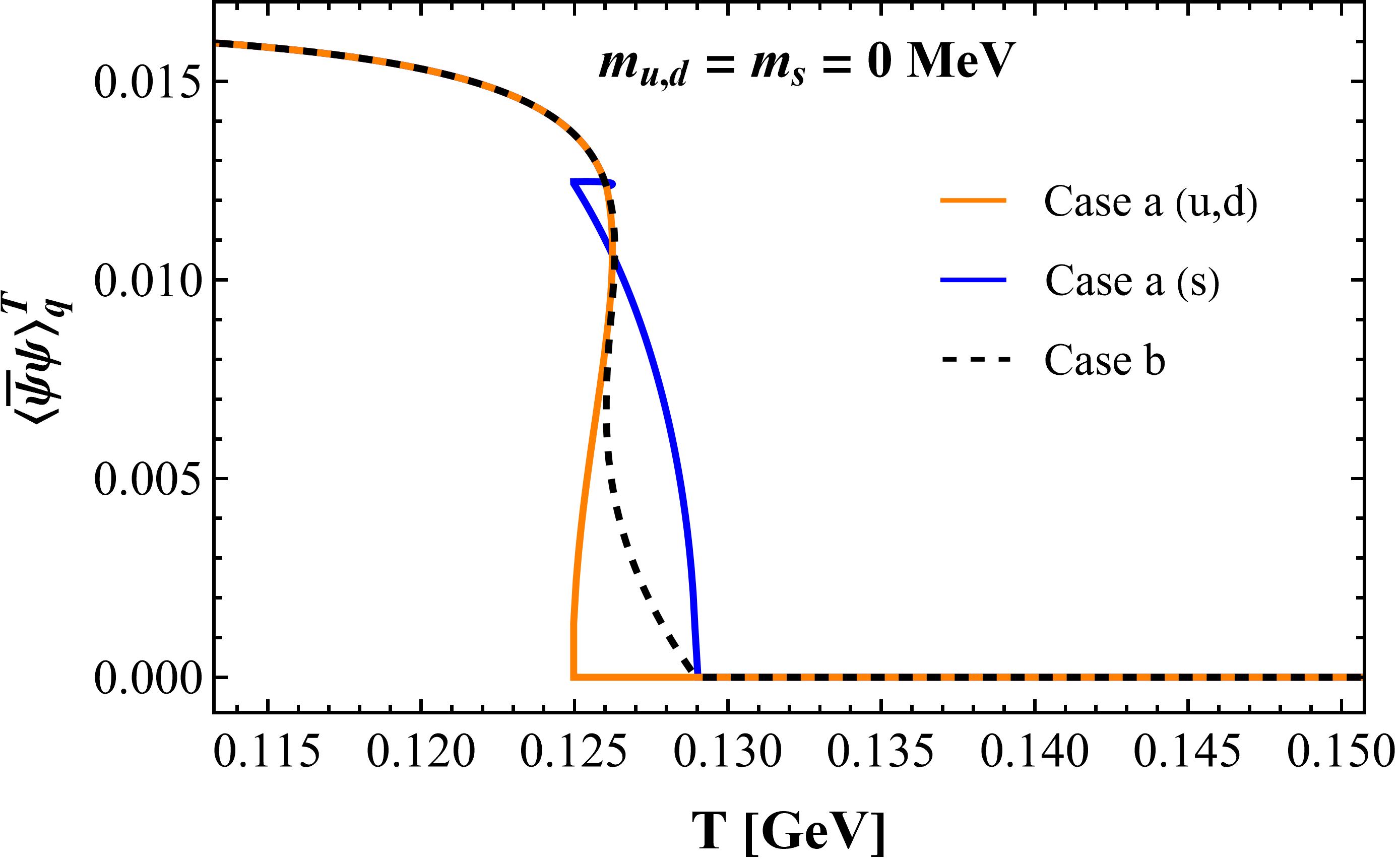}
\includegraphics[width=0.49\linewidth]{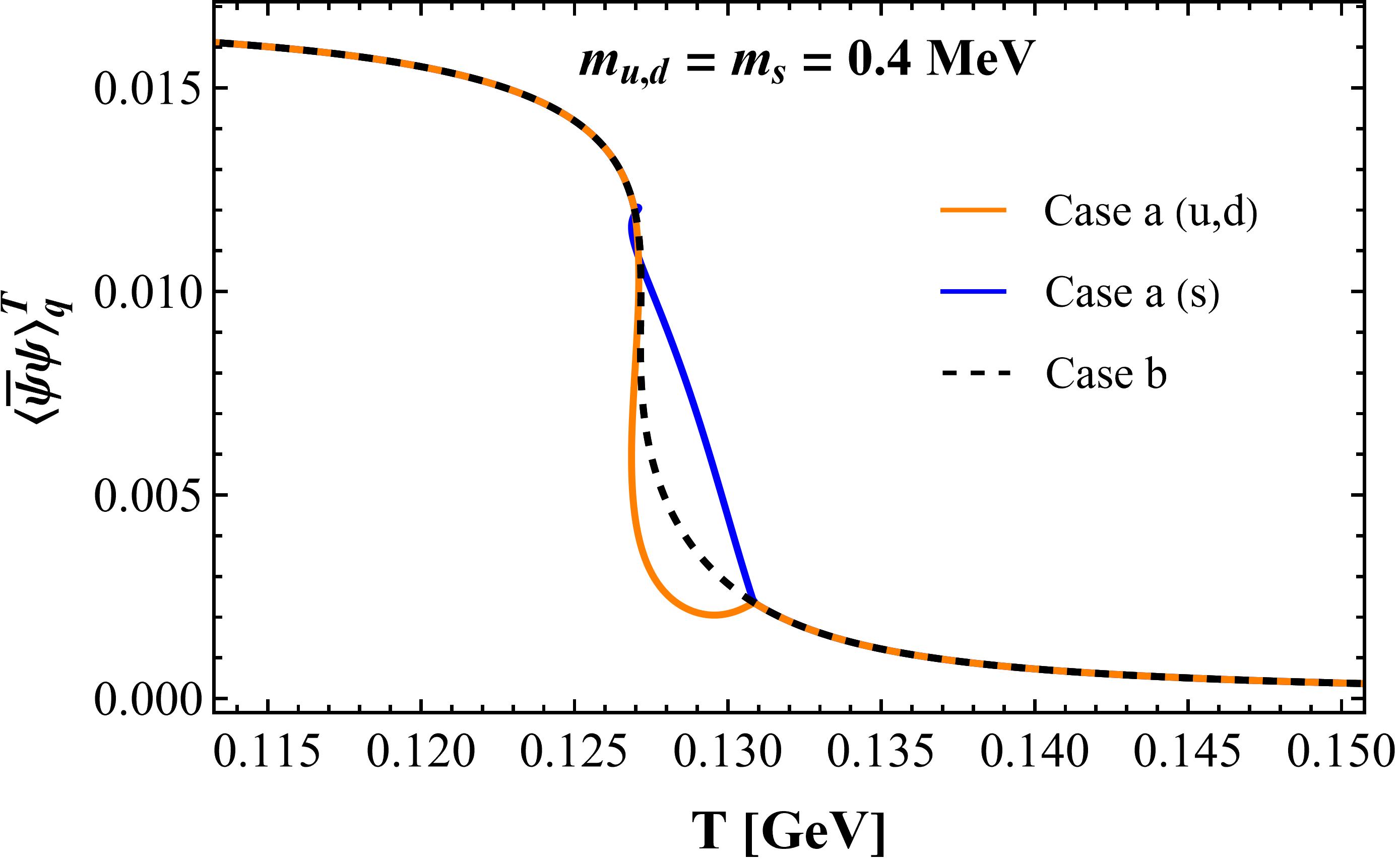}
\includegraphics[width=0.49\linewidth]{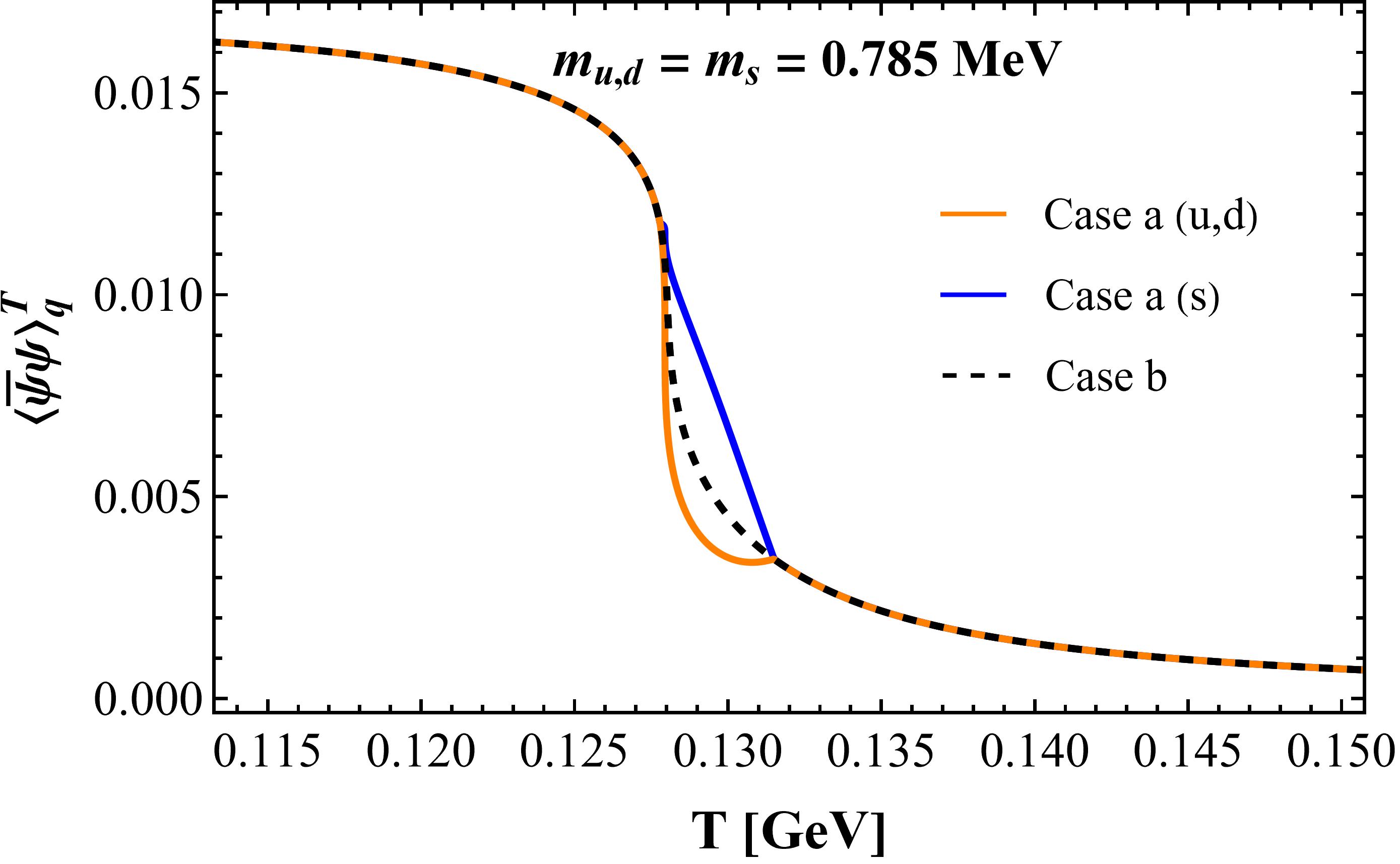}
\includegraphics[width=0.49\linewidth]{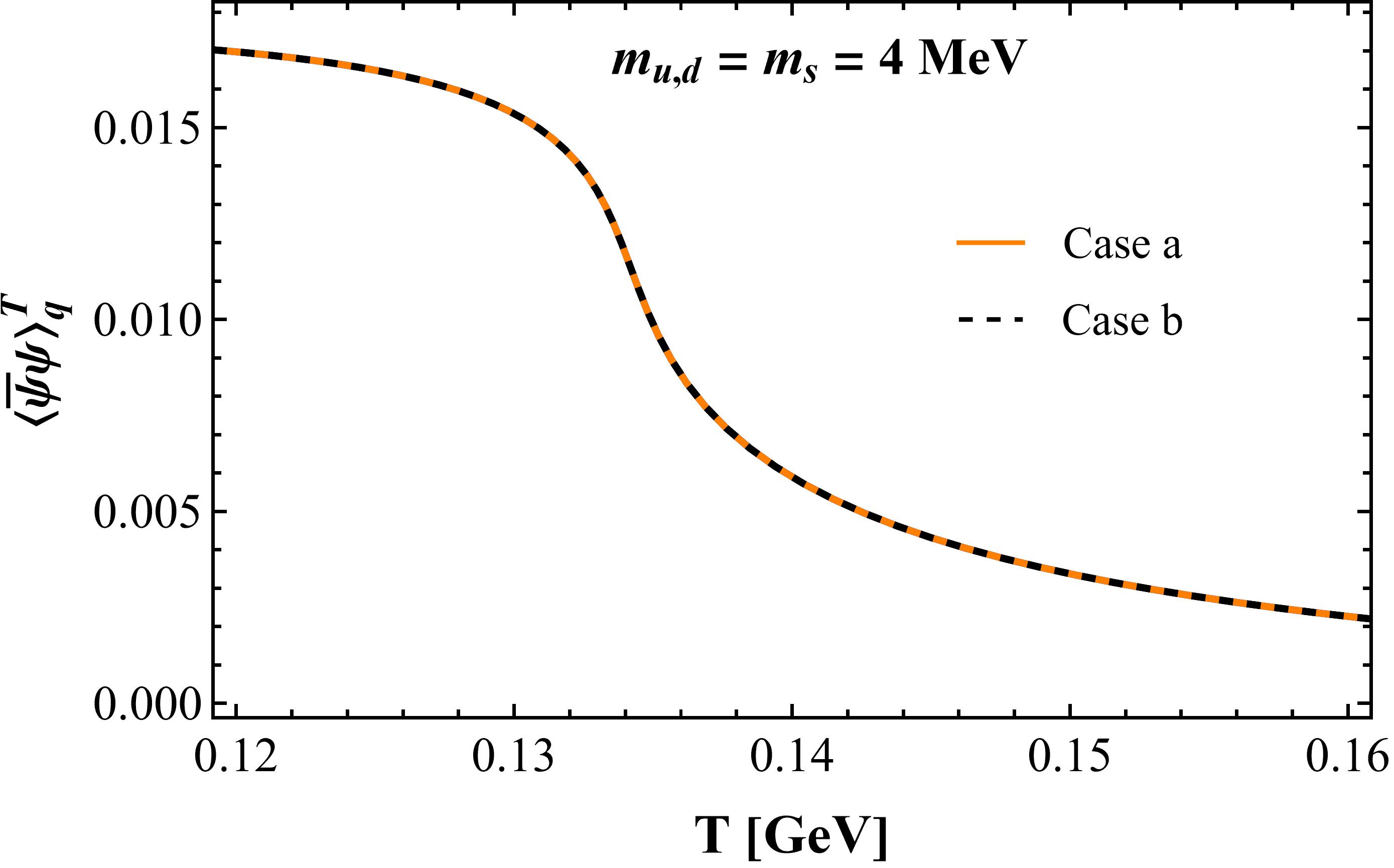}
    \caption{The temperature dependence of the chiral condensates $\langle\bar{\psi}\psi\rangle^T_q$ for quark masses $m_{u,d} = m_s =$ 0, 0.4, 0.785, and 4 MeV in Case a and Case b. In Case a, the orange solid line represents the condensates for light quarks ($u,d$), while the blue solid line corresponds to the strange quark ($s$). In Case b, the black dashed line denotes the condensates for all quark flavors ($u,d,s$).}
    \label{fig:9}
\end{figure}

As illustrated in Fig.~\ref{fig:9}, in Case a, the chiral condensates of the light quark (orange solid line) and the strange quark (blue solid line) show significant differences near the transition temperature, as seen in the upper-left, upper-right, and lower-left panels. These differ markedly from the results of Case b (black dashed line). This splitting behavior appears for very small quark masses. However, as shown in the bottom-right panel of Fig. \ref{fig:9}, when the quark mass increases (e.g., $m_{u,d} = m_s = 4$ MeV), the equations yield a single solution throughout the temperature range, giving $\chi_u = \chi_s$ and identical chiral condensates.

\begin{figure}[htbp]
    \centering   \includegraphics[width=0.49\linewidth]{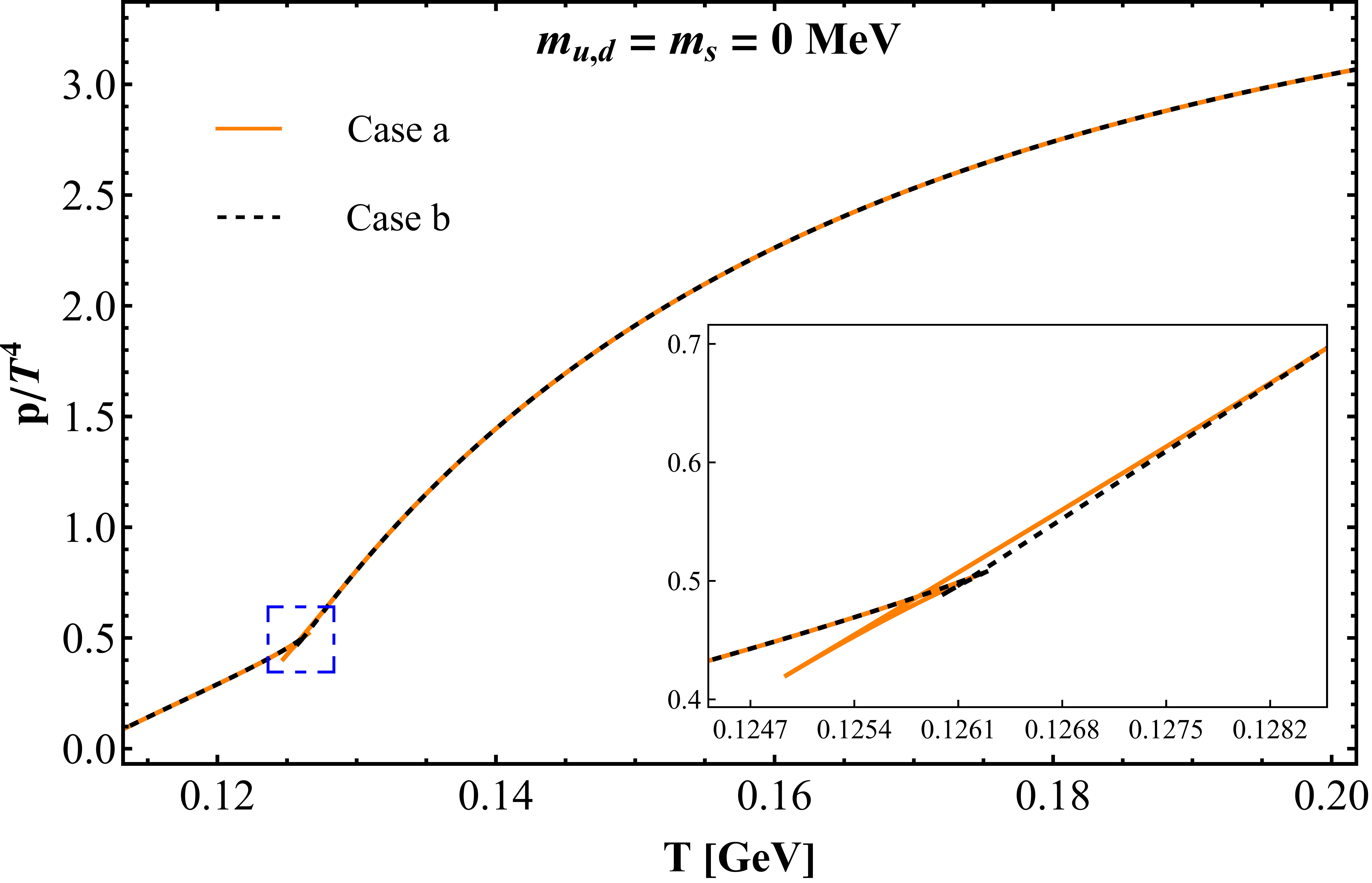}   \includegraphics[width=0.49\linewidth]{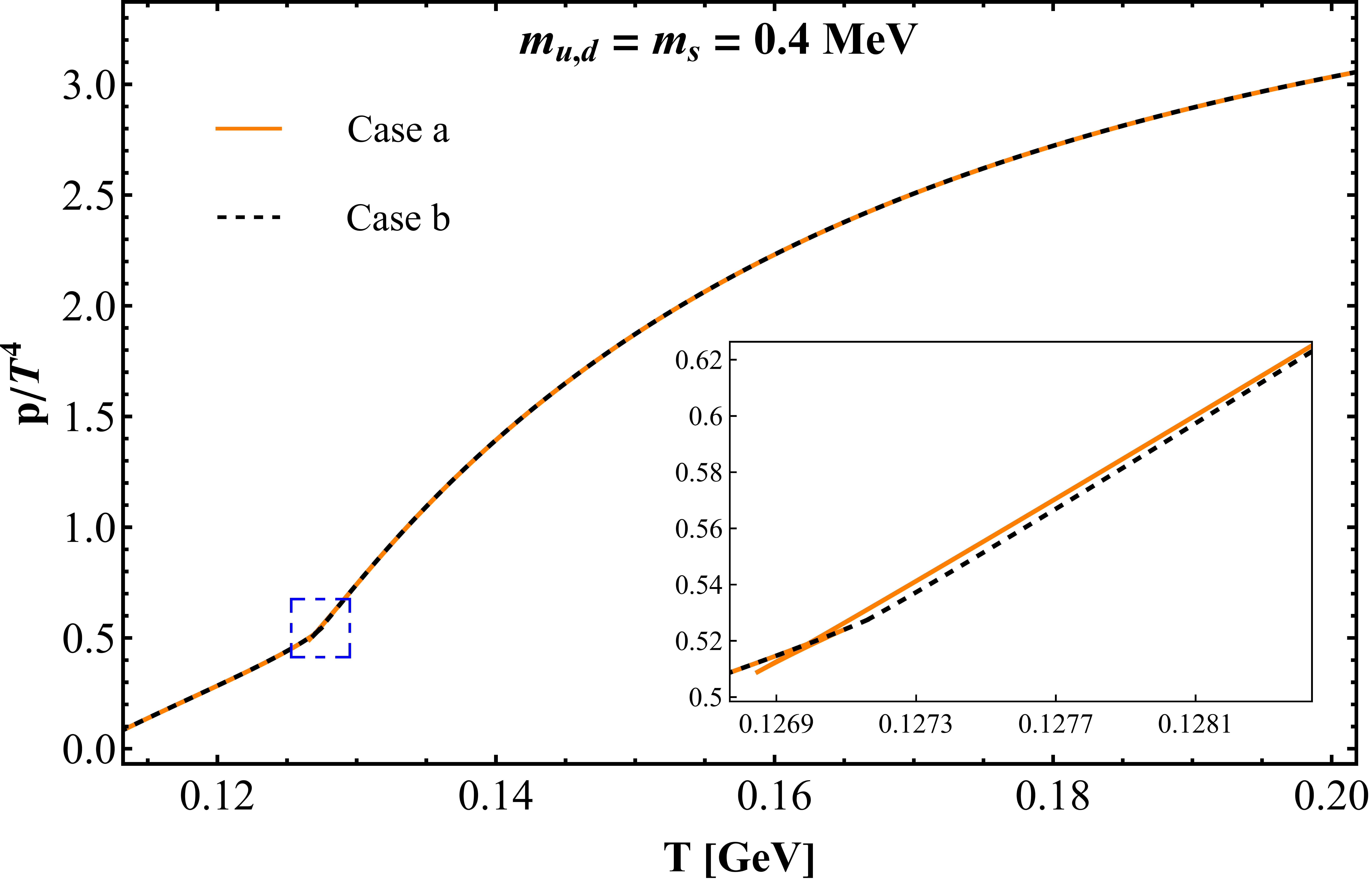}  \includegraphics[width=0.49\linewidth]{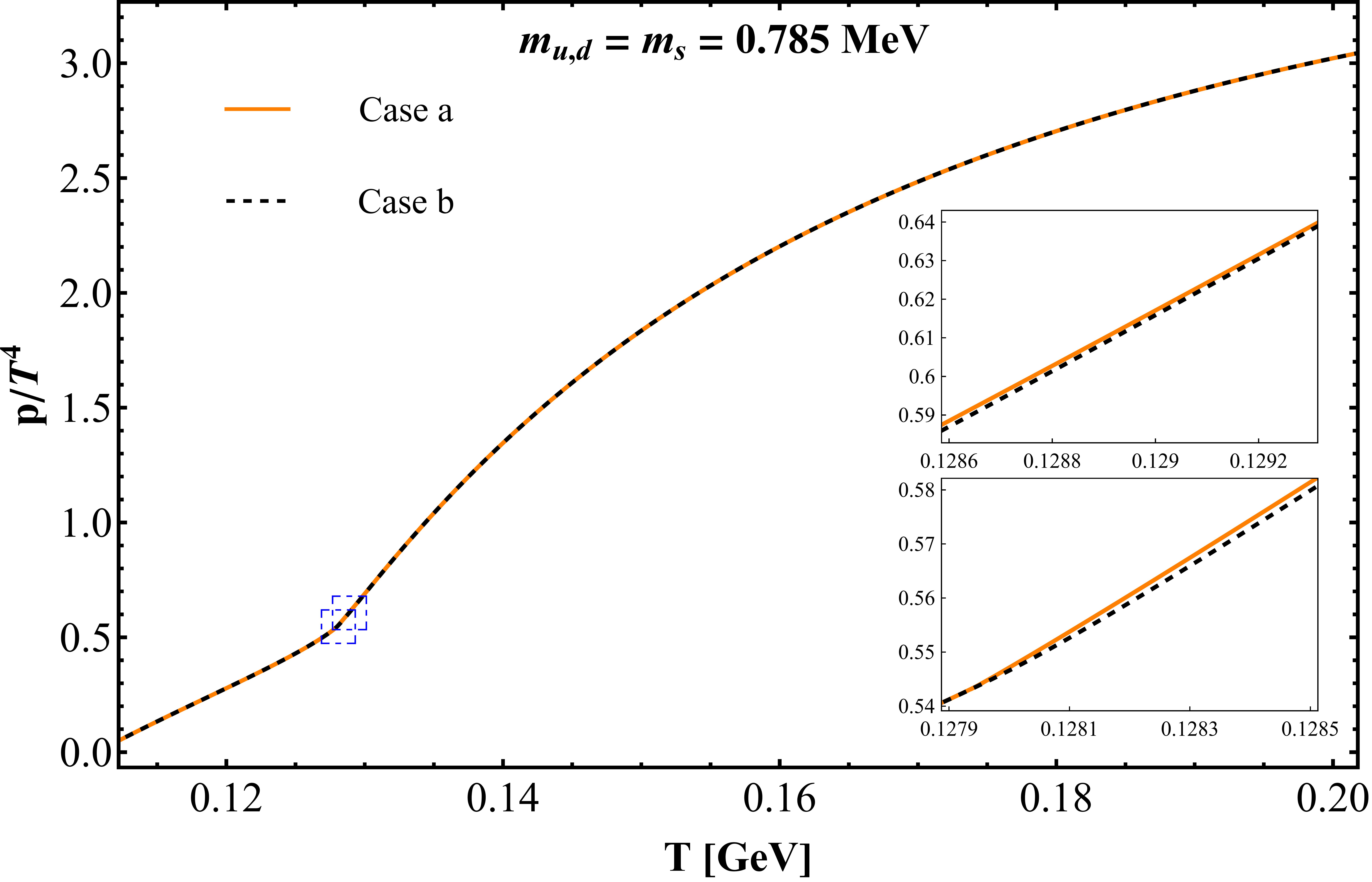}  \includegraphics[width=0.49\linewidth]{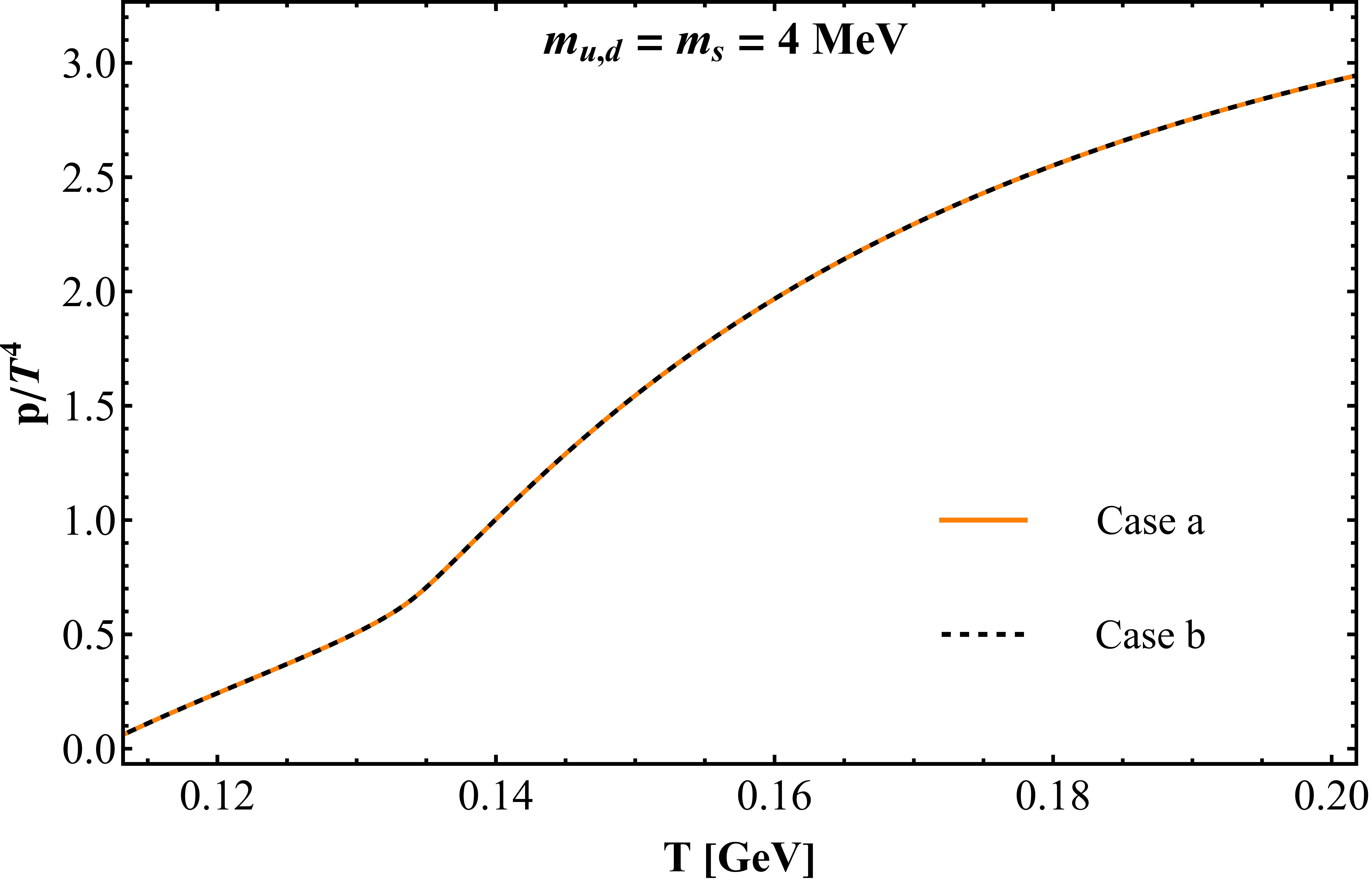} 
    \caption{The temperature dependence of the pressure $p$ for quark masses $m_{u,d} = m_s =$ 0, 0.4, 0.785, and 4 MeV in Case a and Case b.}
    \label{fig:10}
\end{figure}

To identify the thermodynamically preferred branch, we compare the pressures $p$ associated with both cases. As shown in Fig. \ref{fig:10}, for quark masses of 0, 0.4, and 0.785 MeV, the difference between Case a and Case b is small; nevertheless, in a narrow temperature range near the transition, we find that $p_{\text{\tiny Case a}} > p_{\text{\tiny Case b}}$ (and $p_{\text{\tiny Case a}} = p_{\text{\tiny Case b}}$ when $m_{u,d} = m_s = 4$ MeV). Since $p = -F$, where $F$ is the free energy, this indicates $F_{\text{\tiny Case a}} < F_{\text{\tiny Case b}}$, implying that Case a is thermodynamically favored.

\begin{figure}[htbp]
\includegraphics[width=0.49\linewidth]{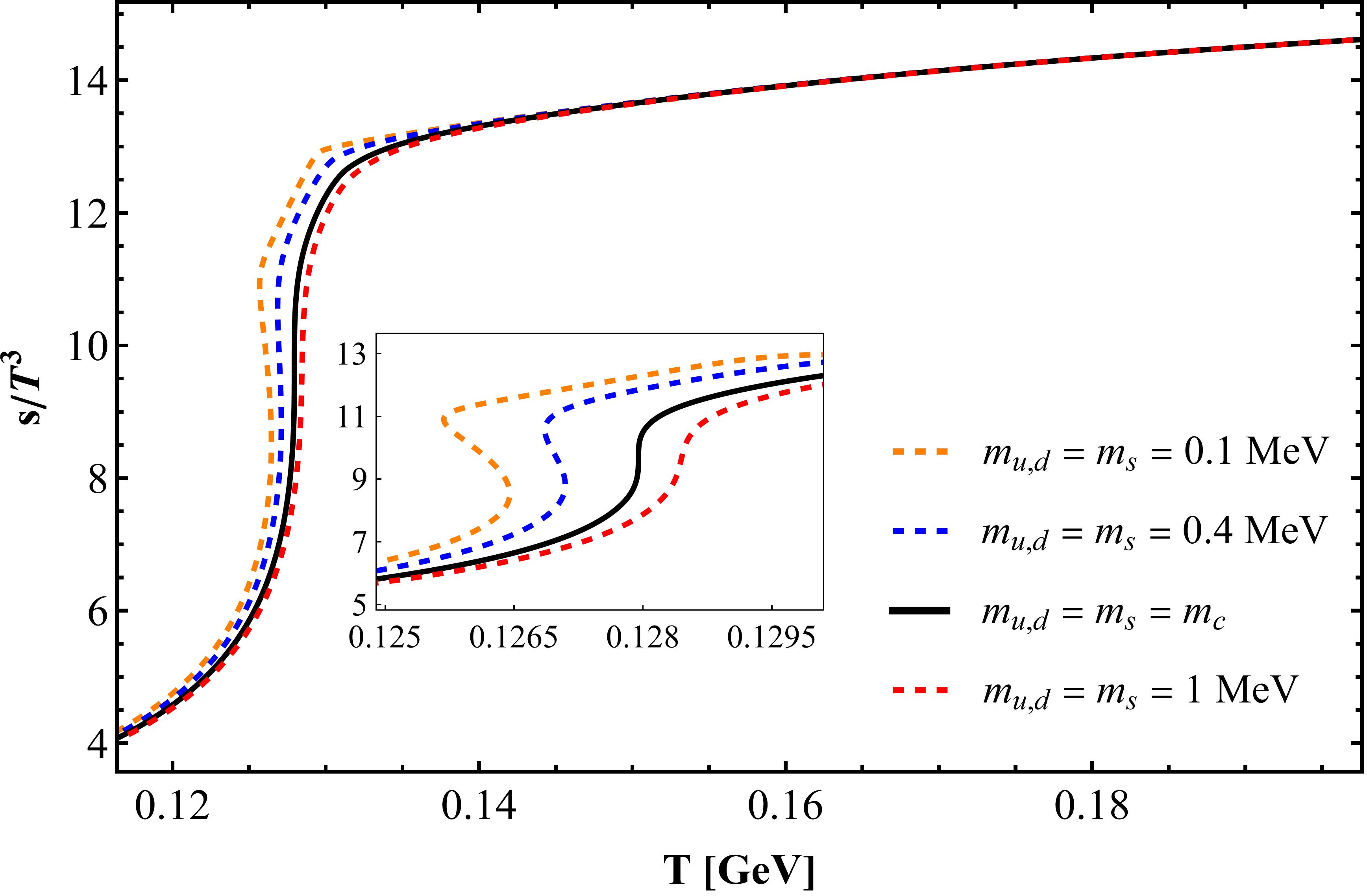}
\includegraphics[width=0.49\linewidth]{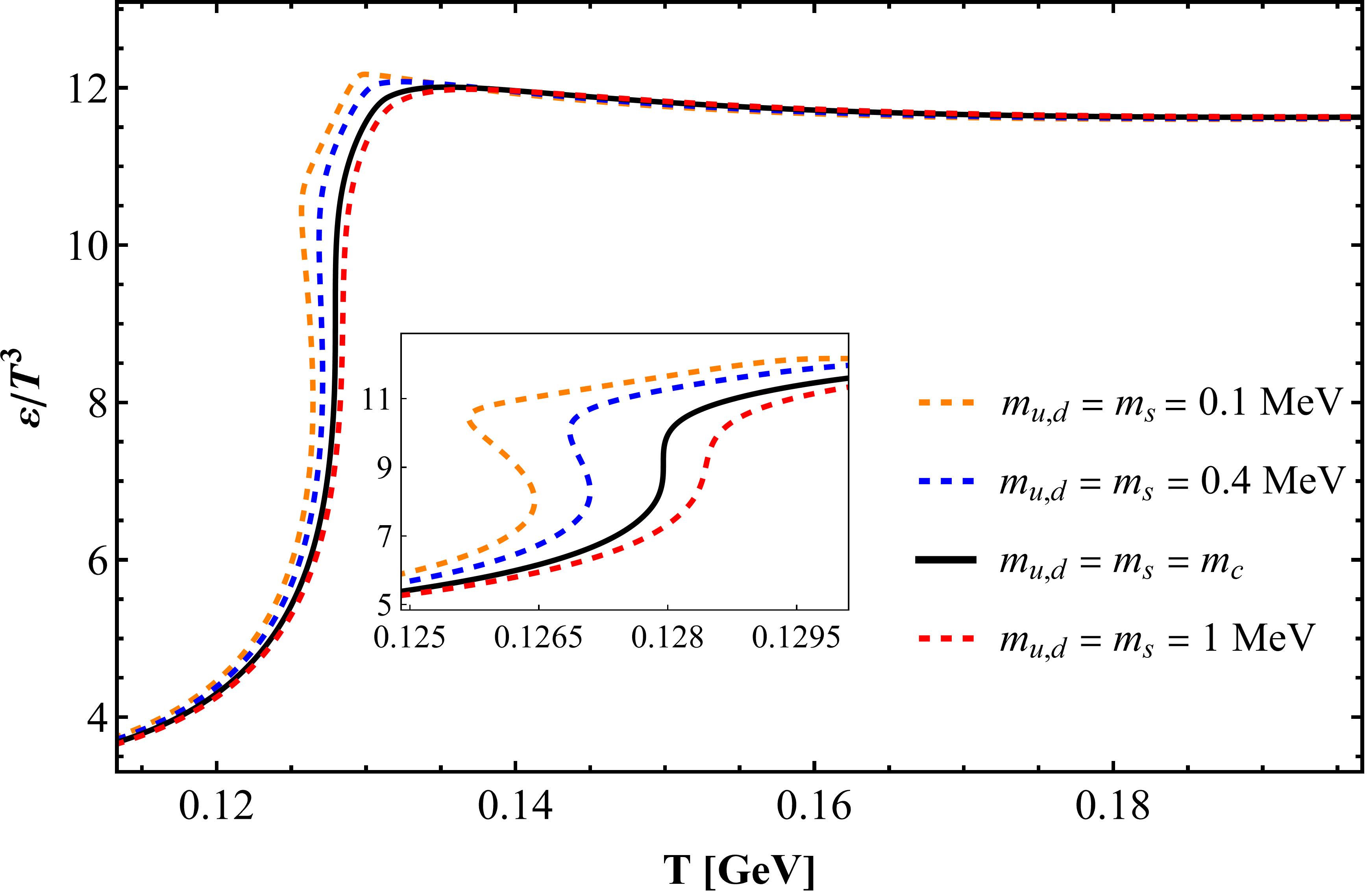}    
    \caption{The temperature dependence of the entropy density $s$ and energy density $\varepsilon$ for quark masses $m_{u,d} = m_s =$ 0.1, 0.4, 0.785, and 1 MeV in Case a, where $m_c = 0.785$ MeV.}
    \label{fig:11}
\end{figure}

Based on the above analysis, we adopt Case a to investigate the flavor-symmetric scenario. By systematically scanning the quark mass and calculating the corresponding chiral condensates and thermodynamic observables, we determine the critical mass $m_c = 0.785$ MeV and mark its position in the quark-mass phase diagram. This value is consistent with lattice estimates locating the boundary at $m_{u,d} = m_s \lesssim m_s^{\mathrm{phy}}/270$ \cite{Endrodi:2007gc,Ding:2011du}.  
The upper panels of Fig. \ref{fig:9} display pronounced inflection points in the transition region, supported by distinct dovetail structures in Fig. \ref{fig:10}, indicating a first-order transition. The lower-left panel of Fig. \ref{fig:9} shows a near-second-order transition, while the lower-right panel depicts a smooth crossover. These results demonstrate that, as the quark mass increases, the chiral phase transition evolves from first-order to second-order, and finally to a crossover — consistent with the quark-mass phase structure.  

In Fig. \ref{fig:11}, we further present the temperature dependence of the entropy density $s$ and energy density $\varepsilon$ for selected quark masses. Notably, the differences in these thermodynamic quantities are most evident at low temperatures. As the temperature increases, both $s$ and $\varepsilon$ converge for different quark masses, consistent with QCD expectations. Moreover, in agreement with lattice QCD studies \cite{Laermann:2003cv}, we find that increasing the quark mass generally raises the critical transition temperature, at least within a finite mass range.

\subsection{The critical exponents}

To gain deeper insight into the universality classes associated with the chiral critical line in our holographic model, we systematically compute the critical exponents along its different segments. As shown in Fig.~\ref{fig:5}, the second-order transition line consists of two distinct segments: a brown segment ($m_{u,d}=0, ~m_s > m_s^{\text{tri}}$) and a blue segment ($m_{u,d}\neq 0, ~m_s < m_s^{\text{tri}}$), which intersect at the tricritical point. These two segments generally fall into different universality classes with the exponents $\beta$ and $\delta$ characterizing the scaling behavior of the order parameter (the chiral condensate $\sigma$) near a critical point:
\begin{align}\label{eq:delta_scaling}
\sigma_q - \sigma_{q,c} \propto \left(1 - \frac{T}{T_c}\right)^\beta, \qquad
\sigma_q - \sigma_{q,c} \propto (m_q - m_{q,c})^{1/\delta},
\end{align}
where $T_c$ and $m_{q,c}$ are the critical temperature and critical quark mass, respectively. In our model, the exponents $(\beta,\delta)$ are extracted numerically from the corresponding logarithmic relations.



\begin{figure}[htbp]
    \centering
    \includegraphics[width=0.472\linewidth]{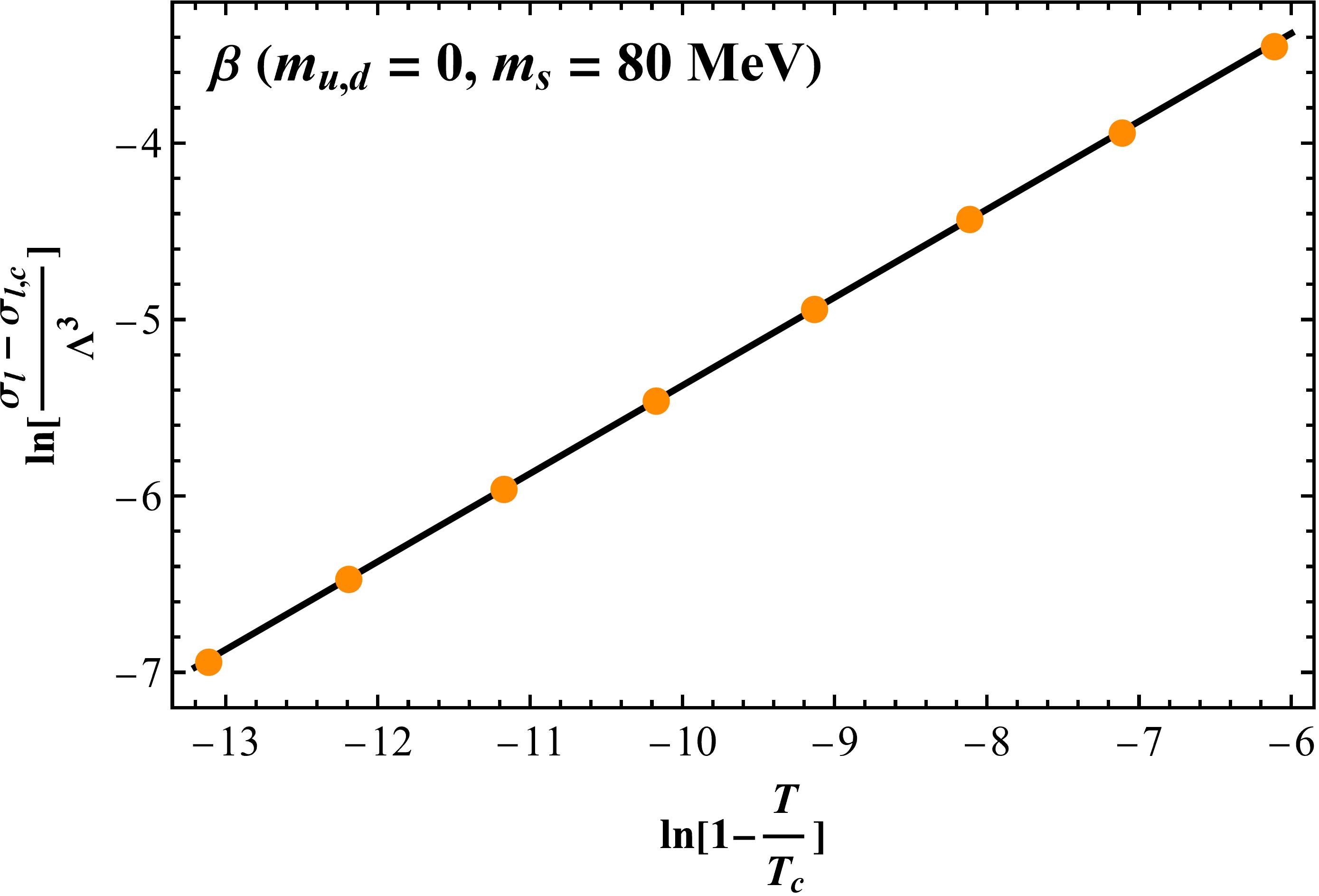}
    \includegraphics[width=0.49\linewidth]{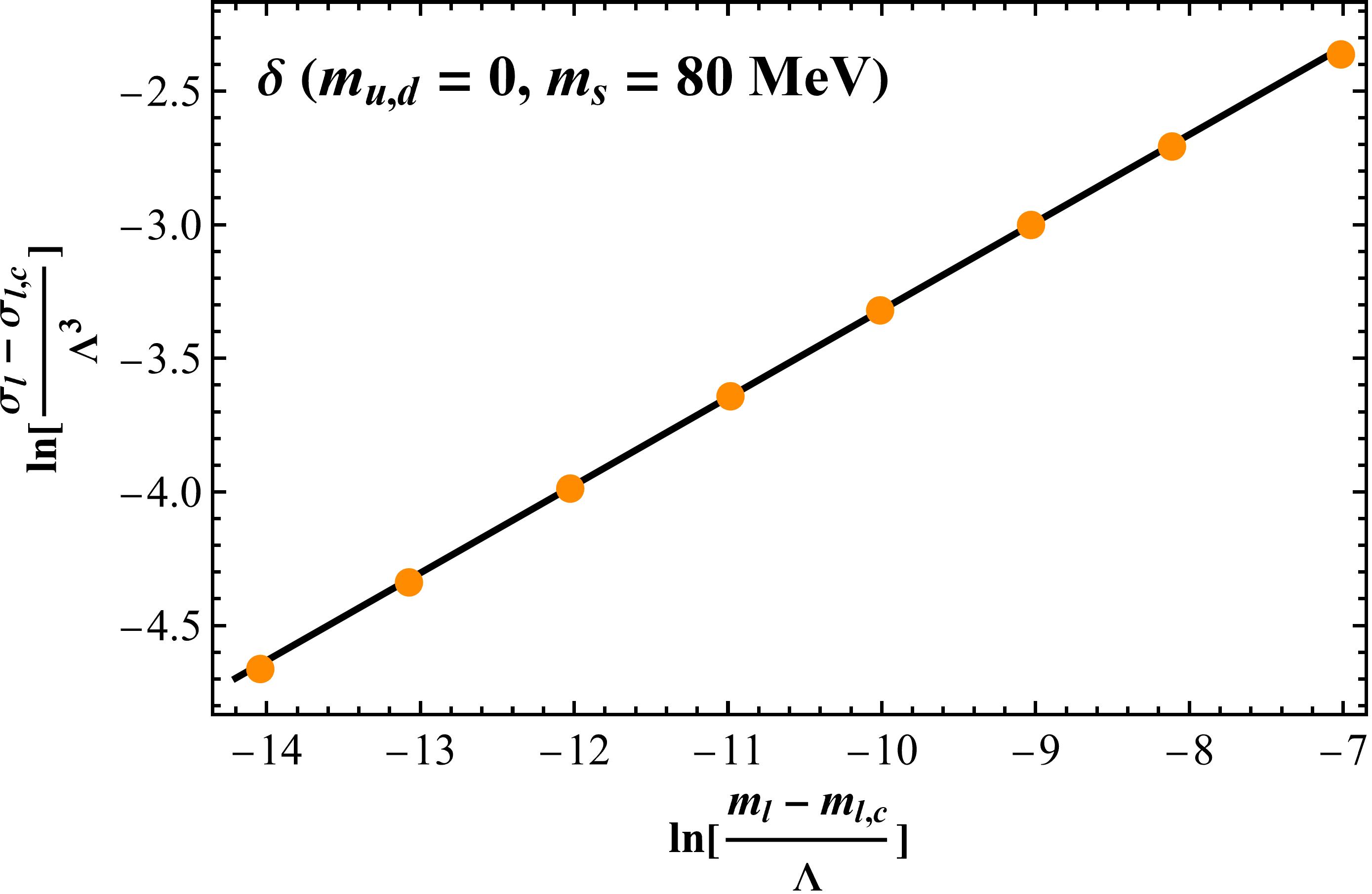}
    \caption{Logarithmic scaling behavior of the light-quark chiral condensate near the critical temperature $T_c$ and critical mass $m_{l,c}$ for $m_{u,d}=0$ and $m_s=80~\mathrm{MeV}$. The orange dots represent the numerical results from our model, while the solid lines denote the best linear fits, given by $y = 0.499\,x - 0.384$ (left) and $y = 0.328\,x - 0.038$ (right).}
    \label{criticalO4}
\end{figure}

\paragraph{Brown segment ($m_{u,d}=0,\, m_s>m_s^{\text{tri}}$).}
For the brown segment of the critical line in the Columbia plot, we select the point $m_{u,d}=0$ and $m_s = 80~\mathrm{MeV}$, corresponding to a critical temperature of $T_{c1}=0.140069701~\mathrm{GeV}$ and a critical light-quark mass $m_{l,c1}=0$. At this temperature, the light-quark condensate takes the critical value $\sigma_{l,c1}=0$. To reduce numerical uncertainties, $T_{c1}$ must be determined with high precision. We then construct
\[
\ln\!\left(\frac{\sigma_l - \sigma_{l,c1}}{\Lambda^3}\right)
\quad \text{vs.} \quad 
\ln\!\left(1 - \frac{T}{T_{c1}}\right),
\qquad
\ln\!\left(\frac{\sigma_l - \sigma_{l,c1}}{\Lambda^3}\right)
\quad \text{vs.} \quad
\ln\!\left(\frac{m_l - m_{l,c1}}{\Lambda}\right),
\]
with $\Lambda = 1~\mathrm{GeV}$ inserted to ensure dimensionless arguments. The scaling curves and fits are shown in Fig.~\ref{criticalO4}, from which we obtain the critical exponents
\[
\beta = 0.499, 
\qquad 
\delta = 3.049.
\]

\begin{figure}[htbp]
    \centering
    \includegraphics[width=0.49\linewidth]{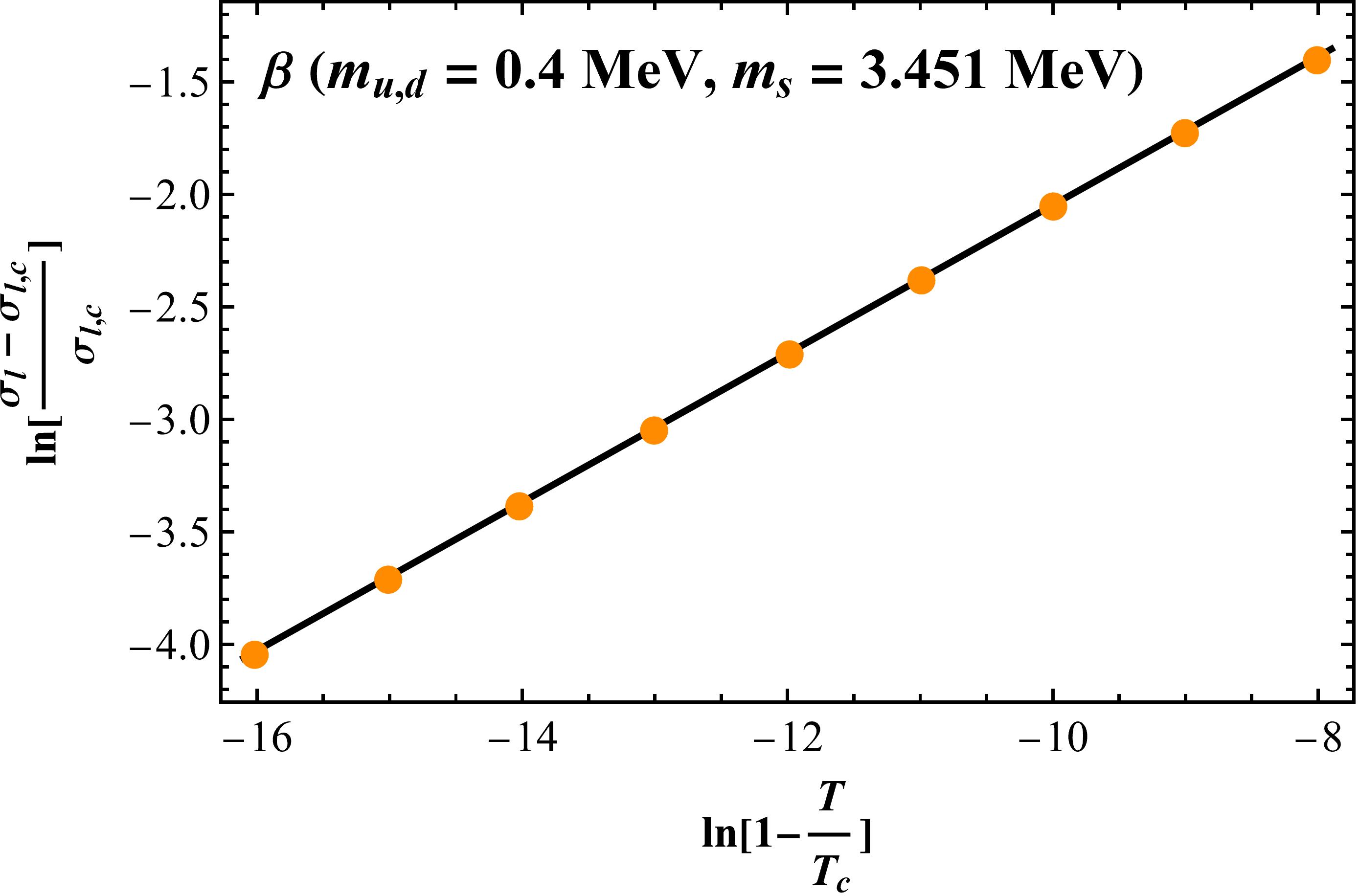}
    \includegraphics[width=0.49\linewidth]{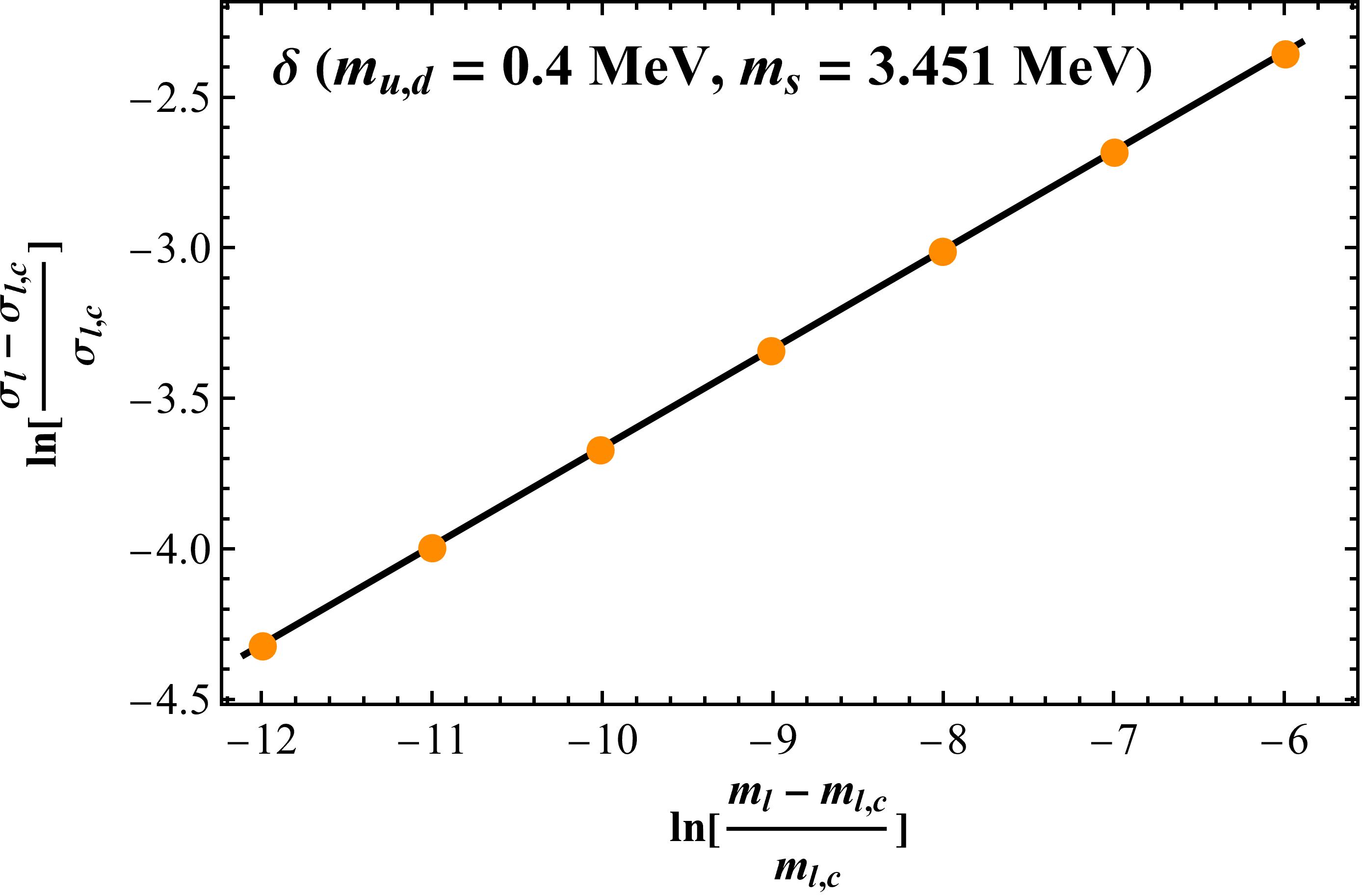}
    \caption{Logarithmic scaling behavior of the light-quark condensate near $T_c$ and $m_{l,c}$ for $m_{u,d} = 0.4~\mathrm{MeV}$ and $m_s = 3.451~\mathrm{MeV}$. The orange dots denote the model results, while the solid lines show the best linear fits, expressed as $y = 0.330\,x + 1.253$ (left) and $y = 0.328\,x - 0.383$ (right).}
    \label{criticalZ2}
\end{figure}

\paragraph{Blue segment ($m_{u,d}\neq 0,\, m_s<m_s^{\text{tri}}$).}
For the blue segment of the critical line, we first consider a representative point at $m_{u,d}=0.4~\mathrm{MeV}$ and $m_s=3.451~\mathrm{MeV}$, for which the critical temperature is $T_{c2}=0.127891727~\mathrm{GeV}$ and the critical quark mass is $m_{l,c2}=0.4~\mathrm{MeV}$. At this temperature, the light-quark condensate takes the value $\sigma_{l,c2}=0.08585~\mathrm{GeV}^3$. To extract the critical exponents, we construct the logarithmic scaling relations
\[
\ln\!\left(\frac{\sigma_l-\sigma_{l,c2}}{\sigma_{l,c2}}\right)
\quad \text{vs.} \quad
\ln\!\left(1-\frac{T}{T_{c2}}\right),
\qquad
\ln\!\left(\frac{\sigma_l-\sigma_{l,c2}}{m_{l,c2}}\right)
\quad \text{vs.} \quad
\ln\!\left(\frac{m_l-m_{l,c2}}{m_{l,c2}}\right),
\]
whose linear fits (Fig.~\ref{criticalZ2}) yield
\[
\beta = 0.330, 
\qquad 
\delta = 3.050.
\]

\begin{figure}[htbp]
    \centering
    \includegraphics[width=0.49\linewidth]{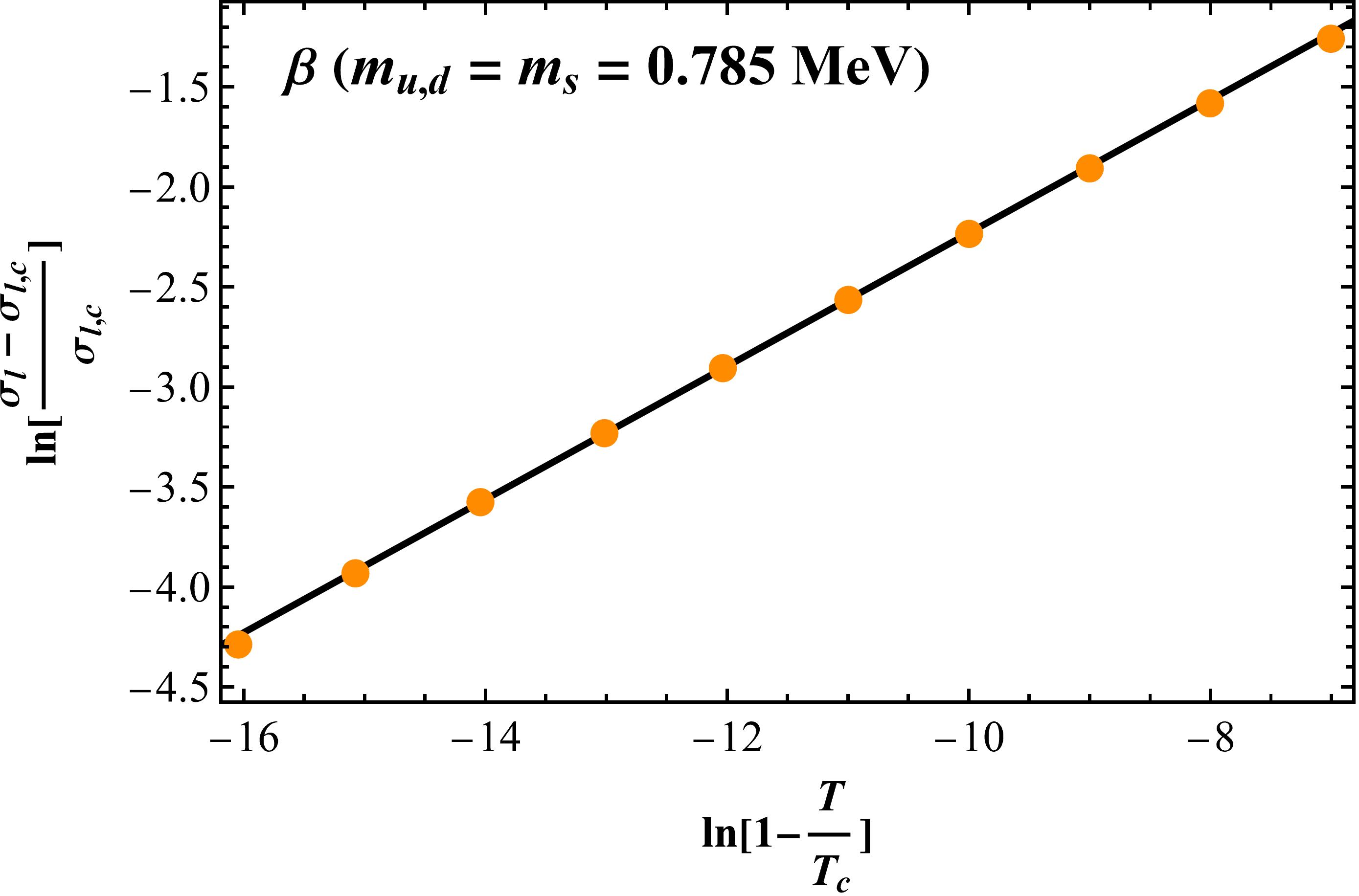}
    \includegraphics[width=0.495\linewidth]{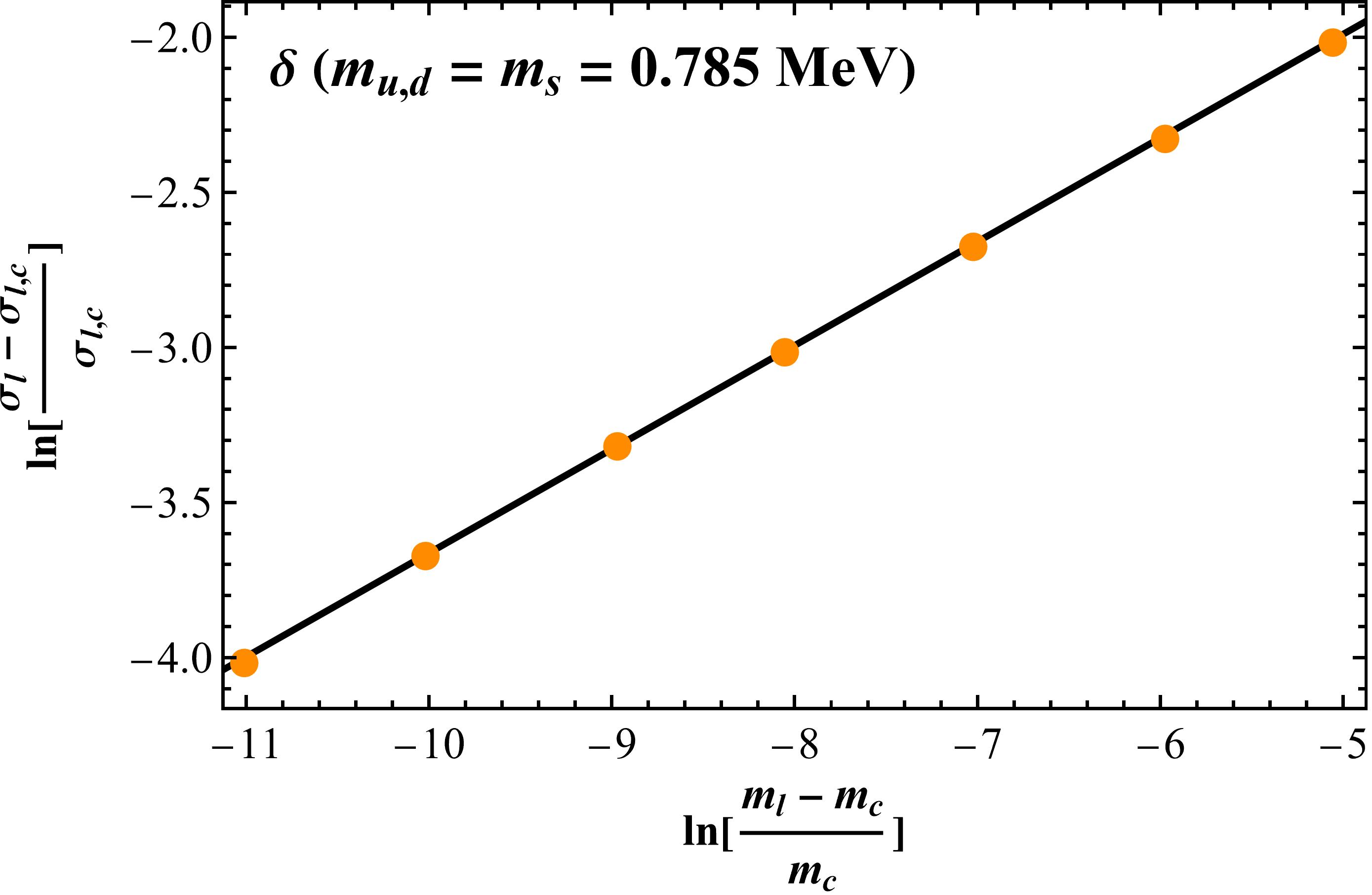}
    \caption{Logarithmic scaling behavior of the light-quark condensate at the flavor-symmetric point $m_{u,d} =m_s =0.785 ~\mathrm{MeV}$. The orange dots represent the model results, while the solid lines denote the best linear fits, given by $y = 0.333\,x + 1.103$ (left) and $y = 0.3346\,x - 0.317$ (right).}
    \label{criticalfsp}
\end{figure}

We also include the flavor-symmetric point $m_{u,d}=m_s=0.785~\mathrm{MeV}$, which lies on the same blue segment of the critical line. Applying the same analysis procedure, with the corresponding critical temperature $T_{c3}=0.127947877~\mathrm{GeV}$, critical mass $m_c=0.785~\mathrm{MeV}$, and critical condensate $\sigma_{l,c3}=0.10871~\mathrm{GeV}^3$, we similarly construct the logarithmic scaling relations. The resulting linear fits (Fig.~\ref{criticalfsp}) give
\[
\beta = 0.333,
\qquad
\delta = 2.989.
\]
These two representative points together characterize the scaling behavior along the entire blue segment.

\begin{figure}[htbp]
    \centering
    \includegraphics[width=0.48\linewidth]{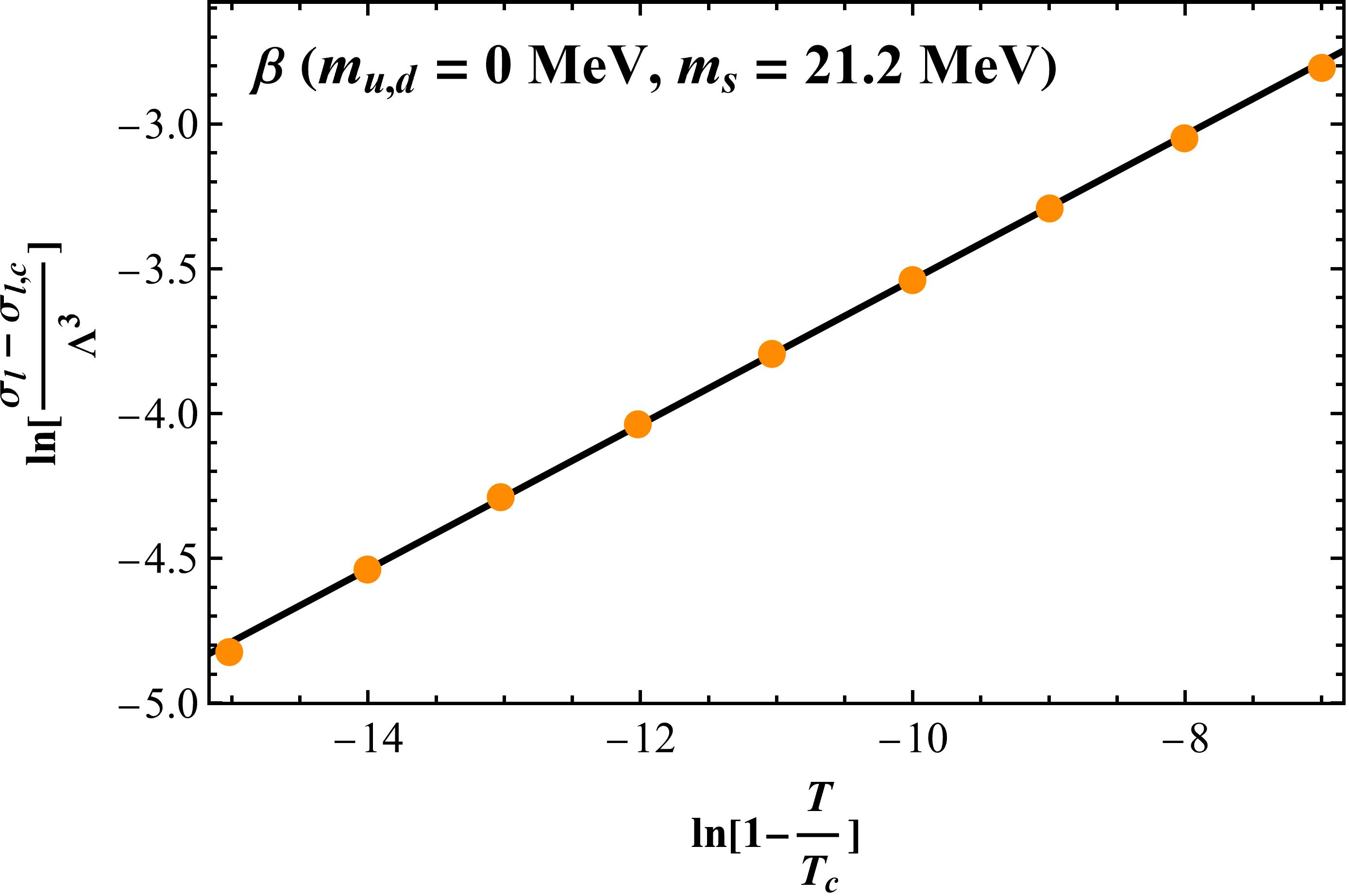}
    \includegraphics[width=0.49\linewidth]{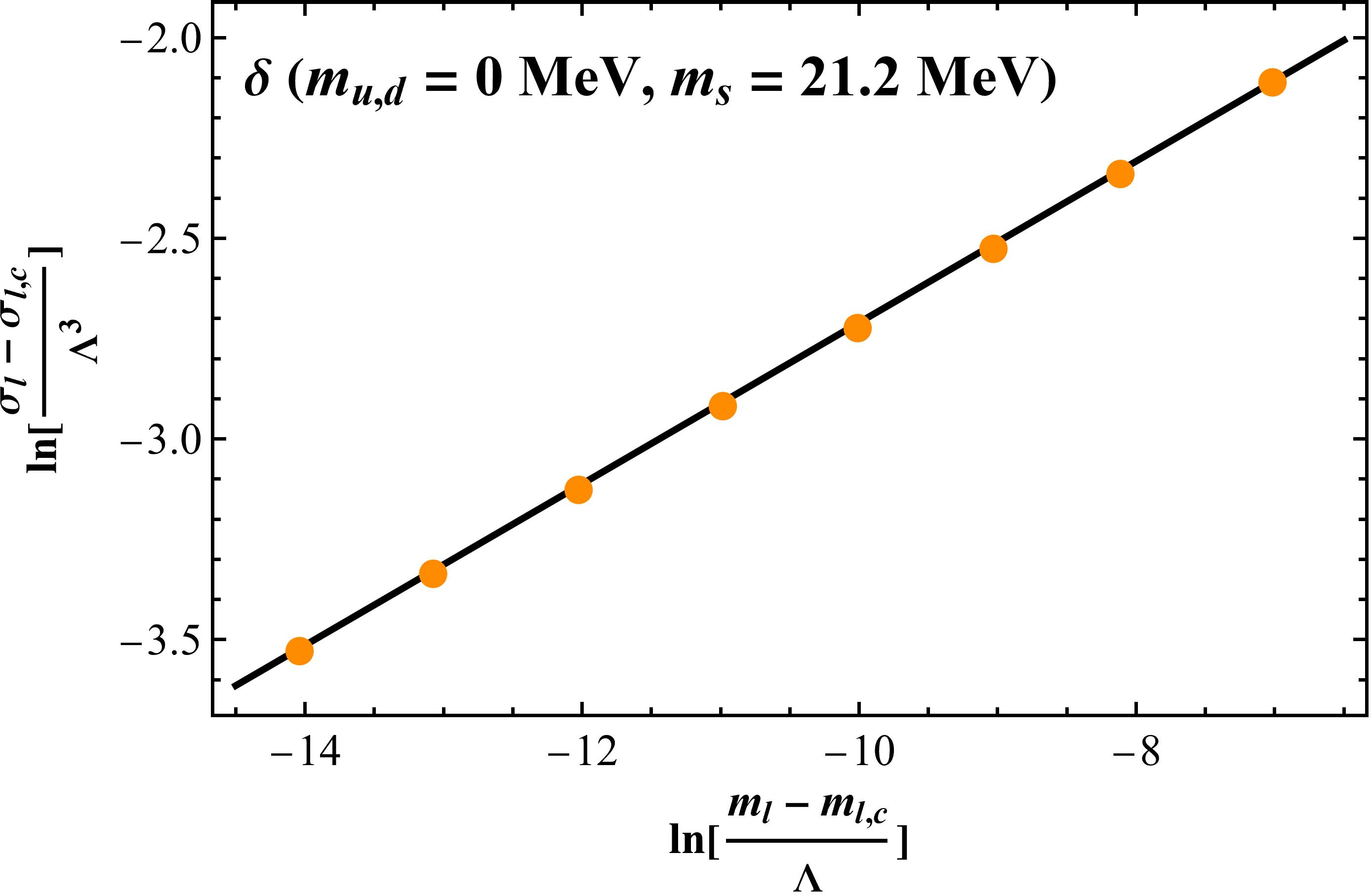}
    \caption{Logarithmic scaling behavior of the light-quark chiral condensate near $T_c$ and $m_{l,c}$ at the tricritical point with $m_{u,d}=0$ and $m_s=21.2~\mathrm{MeV}$. The orange dots indicate the numerical results from the model, while the solid lines correspond to the best linear fits, given by $y = 0.250\,x - 1.038$ (left) and $y = 0.201\,x - 0.700$ (right).}
    \label{criticalTri}
\end{figure}

\paragraph{tricritical point.}
At the tricritical point with $m_{u,d}=0$ and $m_s=21.2~\mathrm{MeV}$, the critical temperature is $T_{c4} = 0.130802993~\mathrm{GeV}$ and the critical light-quark mass is $m_{l,c4} =0$. At this temperature, the condensate satisfies $\sigma_{l,c4} =0$. Using the same procedure as for the brown segment and constructing the logarithmic relations with $\Lambda = 1~\mathrm{GeV}$, we obtain the exponents
\[
\beta = 0.250, 
\qquad 
\delta = 4.975,
\]
as shown in Fig.~\ref{criticalTri}.

As in Ref.~\cite{Chen:2018msc}, the numerical results in our EDF model agree, within numerical uncertainties, with the analytic expectations for the Landau mean-field and tricritical exponents. The brown (two-flavor--like) segment yields mean-field values \((\beta,\delta) \approx (1/2,3)\); the blue (effectively three-flavor--dominated) segment gives \((\beta,\delta) \approx (1/3,3)\); and the tricritical point reproduces the Landau tricritical values \((\beta,\delta) \approx (1/4,5)\). However, these exponents differ markedly from the non-mean-field values expected in QCD, as determined from lattice simulations, functional renormalization group (FRG) analyses, and Dyson--Schwinger equation (DSE) studies \cite{CP-PACS:2000phc,Karsch:2010ya,Kaczmarek:2011zz,Burger:2011zc,Grahl:2014fna,Wang:2015bky,Fischer:2011pk,Fischer:2012vc,Kanaya:1994qe,Engels:2003nq,Bazavov:2017xul,Ding:2017giu}. In QCD, the brown segment ($m_{u,d}\to 0$ with large $m_s$) belongs to the O(4) universality class, characterized by $\beta \approx 0.385$ and $\delta \approx 4.824$, reflecting the $\mathrm{SU}(2)_L \times \mathrm{SU}(2)_R \simeq \mathrm{O}(4)$ symmetry breaking to $\mathrm{O}(3)$. The blue segment (small $m_{u,d,s}$) falls into the Z(2) universality class, where $\beta \approx 0.327$ and $\delta \approx 4.789$.

We further examined the sensitivity of the extracted critical exponents to variations of the flavor-sector parameters in the bulk scalar potential. Varying the quartic coupling \(\lambda\) and the determinant coupling \(\gamma\) over ranges that significantly modify the size of the first-order region shifts the phase-boundary location but leaves the measured critical exponents unchanged within numerical errors, consistent with Ref.~\cite{Chen:2018msc}, although their analysis was performed on a fixed background. Likewise, introducing a \(\chi^6\) term and re-fitting the flavor-sector parameters (while keeping the other sectors fixed) produces no noticeable alteration in the universal exponents. These observations suggest that, in the present setup, the critical behavior is controlled primarily by the structural features of the model---namely symmetry and the leading analytic structure of the bulk equations---rather than by moderate variations of the coupling constants in the scalar potential.

The emergence of mean-field critical behavior in our holographic construction can be traced to several structural aspects of the model. The bottom-up dual is formulated at the level of classical gravity, where fluctuations that normally drive critical exponents away from mean-field values are parametrically suppressed. In the present EDF setup, the dilaton and metric backgrounds are specified phenomenologically and remain effectively non-dynamical near the critical point; consequently, the flavor-scalar sector does not induce sufficiently strong back-reaction to generate nontrivial infrared scaling, leading to critical dynamics governed by a Landau-type effective theory. Furthermore, the bulk scalar potential contains only a limited set of analytic (typically polynomial) terms, so that the near-critical expansion reduces to algebraic relations whose leading-order balance yields Landau mean-field exponents. This explains why even the inclusion of moderate higher-order interactions, such as a $\chi^6$ term, does not qualitatively modify the resulting critical behavior.

\section{Summary and discussion}\label{sec:5}

In this work we developed a 2+1--flavor Einstein--dilaton--flavor holographic model and calibrated it using lattice-QCD data for the equation of state and temperature-dependent chiral condensates. By integrating numerical solvers with machine-learning optimization methods, we efficiently explored the high-dimensional parameter space and obtained a single parameter set that reproduces lattice thermodynamics and chiral observables with good accuracy. The resulting description of the crossover at the physical point is consistent with HotQCD and W-B results, demonstrating that machine learning can substantially streamline and stabilize holographic model calibration.

With the model fixed, we mapped the quark-mass phase diagram and determined the structure of the Columbia plot across a wide range of light and strange quark masses. The resulting phase diagram exhibits the characteristic structure expected for 2+1 flavors, featuring a small first-order region bounded by a second-order critical line and terminating at a tricritical point on the $m_{u,d}=0$ axis. The extracted benchmark masses including the flavor-symmetric critical mass and the tricritical strange mass indicate that the first-order region is small and that the physical point lies firmly in the crossover regime, consistent with current lattice constraints \cite{Cuteri:2017zcb,Bernhardt:2023hpr,Guenther:2020jwe}. The model thus provides a coherent holographic picture of the 2+1--flavor thermal phase structure.

We also examined the critical exponents along the second-order critical line and at the tricritical point. While our model successfully reproduces the qualitative structure of the Columbia plot and identifies the expected universality classes, the numerical values of the critical exponents along the chiral critical line remain governed by mean-field dynamics in the present formulation. To achieve the nontrivial exponents expected in QCD, we should go beyond the present classical treatment by incorporating fluctuation effects. Moreover, developing fully dynamical backreaction schemes in which the flavor scalar, dilaton, and bulk geometry mutually respond in the near-critical infrared (rather than only through coupled classical backgrounds) may enable the emergence of non-mean-field universality classes such as O(4) or Z(2).


Future work should pursue several complementary directions to improve the predictive capacity of the EDF framework for the QCD phase structure. First, extending the analysis to finite baryon chemical potential will enable the search and characterization of a possible critical endpoint and clarify how the Columbia-plot structure evolves at $\mu > 0$. Machine-learning tools are expected to be particularly valuable as the parameter space and computational cost increase significantly. A preliminary study along this direction has been carried out in the short companion work \cite{Shen:2025zkj}. Second, systematically exploring model dependence by varying the dilaton potential, anomaly terms (such as the 't~Hooft determinant), and higher-order scalar interactions, while statistically calibrating against a broader set of lattice observables will help distinguish robust predictions from artifacts of specific model choices. Finally, investigating real-time dynamics and transport properties, including spectral functions, condensate relaxation, and critical slowing down, would bridge equilibrium critical behavior with experimentally accessible signatures and provide further constraints on holographic constructions.

\section*{Acknowledgments}
We are grateful to Danning Li, Xun Chen, Hong-An Zeng, and Ling-Jun Guo for their valuable discussions. This work is supported by Hunan Provincial Natural Science Foundation of China (Grants No. 2023JJ30115 and No. 2024JJ3004), and also supported by the National Key Research and Development Program of China under Grant No. 2020YFC2201501 and the National Science Foundation of China (NSFC) under Grants No.~12347103 and No.~11821505.

\appendix

\section{Formulation of the EDF System}\label{app1}

This section outlines the construction of the action~(\ref{Stotal}) for the EDF system. The metric ansatz in the string frame is
\begin{equation}\label{stringfmetric} 
    ds^2 = \frac{L^2 e^{2 A_S(z)}}{z^2} \left(-f(z)\, dt^2 + \frac{dz^2}{f(z)} + dx^i dx^i\right),
\end{equation}
where $i = 1,2,3$, and $L = 1$ denotes the curvature radius of the asymptotic AdS$_5$ spacetime. The coordinate $z$ represents the holographic radial direction.

The total action of the 2+1-flavor holographic QCD model consists of two parts, $S = S_G + S_M$. The term $S_G$ describes the gravitational background and is given by the Einstein--dilaton action in the string frame:
\begin{align}\label{2+1nonmi-gra-str}
    S_G = \frac{1}{2\kappa_N^2} \int d^5x\, \sqrt{-g_S}\, e^{-2\phi}
    \left[ R_S + 4\,\partial_M \phi\, \partial^M \phi - V(\phi) \right],
\end{align}
where $\kappa_N = \sqrt{8\pi G_5}$. A suitably chosen dilaton potential $V(\phi)$ is required to reproduce QCD-like thermodynamic behavior; its precise form will be discussed later.

The flavor-sector action differs from that of the two-flavor case \cite{Liu:2023pbt}. To correctly capture the chiral phase transition in the 2+1-flavor system, an additional term proportional to the ’t~Hooft determinant $\det[X]$ must be included, as shown in Refs.~\cite{Chelabi:2015cwn,Chelabi:2015gpc,Fang:2018vkp}. The matter action $S_M$ of the improved soft-wall model is
\begin{equation}\label{SM1}
    \begin{split}
        S_M = -\kappa \int d^5x\, \sqrt{-g_S}\, e^{-\phi(z)}
        \Big\lbrace \mathrm{Tr}\!\Big[ |DX|^2 + V_X(X,\Phi)
        + \frac{1}{4g_5^2} (F_L^2 + F_R^2) \Big] 
        + \gamma\, \det|X| \Big\rbrace,
    \end{split}
\end{equation}
where $D^M X = \partial^M X - i A_L^M X + i X A_R^M$ is the covariant derivative, and the gauge-field strengths are
$F^{MN}_{L,R} = \partial^M A^{N}_{L,R} - \partial^N A^{M}_{L,R} - i[A^M_{L,R},A^N_{L,R}]$.
In what follows, the gauge fields $A_L$ and $A_R$ are treated as vacuum fluctuations and are neglected. The potential for the bulk scalar $X$ and dilaton $\Phi$ is
\begin{align}\label{2+1VX}
    V_X(X,\Phi) = \left(m_5^2 - \Phi(\phi)\right) |X|^2 + \lambda |X|^4,
\end{align}
where the bulk mass satisfies $m_5^2 L^2 = \Delta_X(\Delta_X - 4) = -3$ for $\Delta_X = 3$, corresponding to the operator $\bar q_R q_L$ dual to $X$ at the boundary~\cite{Erlich:2005qh}. The vacuum expectation value (VEV) of the scalar field $X$ in the 2+1-flavor case is taken as
\begin{align}\label{2+1X}
    \langle X \rangle = \frac{1}{\sqrt{2}}
    \begin{pmatrix}
        \chi_u(z) & 0 & 0 \\
        0 & \chi_d(z) & 0 \\
        0 & 0 & \chi_s(z)
    \end{pmatrix},
\end{align}
with $\chi_u = \chi_d$. The chiral condensates are encoded in the UV expansions of $\chi_u$ and $\chi_s$.

To analyze the chiral phase transition, we retain only the VEVs of the matter fields in the action~(\ref{SM1}), reducing the holographic QCD model to the EDF system:
\begin{align}\label{S1}
    S &= S_G + S_\chi   \notag\\
    &= \frac{1}{2 \kappa_N^2} \int d^5x \sqrt{-g_S}\, e^{-2\phi}\left[R_S + 4(\partial \phi)^2 -V(\phi)\right.   \notag\\
    &\quad\left. -\beta e^{\phi} \left((\partial\chi_u)^2+\frac{1}{2}(\partial\chi_s)^2 +V_\chi(\chi_u,\chi_s,\phi)\right) \right],
\end{align}
where $\beta = 16\pi G_5\, \kappa$ characterizes the coupling between the background geometry and the flavor fields. The effective potential $V_\chi$ for $\chi_u$, $\chi_s$, and $\phi$ is
\begin{align}
    V_\chi(\chi_u,\chi_s,\phi)
    &= \mathrm{Tr}\!\left[V_X(X,\Phi)\right] + \gamma\, \det|X| \notag\\
    &= -\frac{1}{2}(3+\Phi(\phi))(2\chi_u^2 + \chi_s^2)
    + \frac{\gamma}{2\sqrt{2}}\, \chi_u^2 \chi_s
    + \frac{\lambda}{4}(2\chi_u^4 + \chi_s^4).
\end{align}

We now transform to the Einstein frame, where the warp factor satisfies $A_E = A_S - \frac{2}{3}\phi$. In this frame, the action~(\ref{S1}) takes the form of Eq.~(\ref{Stotal}), with
\begin{equation}
    \begin{split}
        V_E(\phi) &= e^{\frac{4\phi}{3}}\, V(\phi), \\
        V(\chi_u,\chi_s,\phi) &= e^{\frac{4\phi}{3}}\, V_\chi(\chi_u,\chi_s,\phi).
    \end{split}
\end{equation}

\section{Coefficients of the UV Expansion for Bulk Fields}\label{app2}

Given the complexity and length of the UV expansion coefficients, we omit them from the main text. For completeness, we list the full set of coefficients below.

Coefficients for $A_E(z)$:
\begin{align}
    a_3 &= \frac{1}{24} \beta  \zeta ^2 (m_u^2 (\sqrt{2} \gamma \zeta m_s -2 p_1 (2 d_1 +11)) -m_s^2 p_1 (2 d_1+11)),  
    \\
    a_4 &= \frac{1}{9331200}\left(1024 p_1^4 \left(756 b_4 -378 \gamma_1^4+73\right)-96 \beta  \zeta^2 p_1^2 \left(3807 d_1^2 +42633 d_1 +567 d_2\right.\right.  \notag\\
        &\quad \left.\left. +115543\right) \left(2 m_u^2 + m_s^2\right) -1296 p_1 (960 p_3-\sqrt{2} \beta \zeta^3 m_u^2 m_s \gamma (423 d_1 +2344))\right.  \notag\\
        &\quad \left. -9 \beta  \left(4 \zeta^4 m_u^4 \left(27 \left(47 \gamma ^2-56  \lambda\right) +\beta  \left(3807 d_1^2 +47304 d_1+145604\right)\right)\right.\right.  \notag\\
        &\quad \left.\left. +4 \zeta ^4 m_u^2 m_s^2 \left(2538 \gamma^2 +\beta  \left(3807 d_1^2+47304 d_1  +145604\right)\right) +\zeta^4 m_s^4 \left(\beta  \left(3807 d_1^2  \right.\right.\right.\right.  \notag\\
        &\quad \left. \left. \left.\left.   +47304 d_1  +145604\right) -3024 \lambda\right) +103680 m_u \sigma_u+51840 m_s \sigma_s\right)\right),   
        \\
    a_{4l} &= \frac{1}{17280} \big( -1024 p_1^4 \left(12 b_4 -6 \gamma_1^4 +1\right) +96 \beta \zeta^2 p_1^2 \left(2 m_u^2 + m_s^2\right) \left(9 d_1^2 +111 d_1 +9 d_2 \right.  \notag\\
        &\quad \left. +286\right) +9 \beta \zeta^4 \left(4 m_u^4 \left(3 \left(\gamma^2 -8 \lambda\right) +\beta \left(9 d_1^2 +108 d_1  +308\right)\right) + 4 m_u^2 m_s^2 \left(6\gamma^2 \right.\right.   \notag\\
        &\quad \left.\left. +\beta \left(9 d_1^2  +108 d_1  +308\right)\right)+m_s^4 \left(\beta  \left(9 d_1^2 +108 d_1 +308\right) -48\lambda \right)\right)   \notag\\
        &\quad -144 \sqrt{2} \beta  \zeta^3 m_u^2 m_s p_1 \gamma (9 d_1 +52)\big).
\end{align}
Coefficients for $\phi(z)$:
\begin{align}
    p_{3l} &= \frac{1}{576} \big(256 p_1^3 \left(12 b_4 -6 \gamma_1^4 +1\right) -12 \beta \zeta^2 p_1 \left(9 d_1^2 +111 d_1 + 9 d_2 +286\right) \left(2 m_u^2+m_s^2\right)   \notag\\
    &\quad +9 \sqrt{2} \beta \zeta^3 m_u^2 m_s (\gamma (9 d_1 +52))\big),  
        \\
    p_4 &= -\frac{1}{2304}\beta  \left(-4 \zeta^2 m_u^2 \big(-4 p_1^2 \left(-576 b_4 (d_1 +6)  +812 d_1^2 +288 \gamma_1^4 (d_1 +6)+96 d_1 d_2 \right.\right.   \notag\\
    &\quad \left.\left. +42 d_1^3 +4515 d_1 +600 d_2 +7064\right) -3 \zeta ^2 m_s^2 \left(3\beta \left(4 d_1^3 + 81 d_1^2 + 6 d_1 d_2 +486 d_1 \right.\right. \right.   \notag\\
    &\quad \left.\left.\left. + 36 d_2 + 872\right) +\gamma^2  (11 d_1 +61)\right) +\sqrt{2} \zeta m_s p_1 \gamma  \left(162 d_1^2  +1947 d_1 +108 d_2 +5156\right)\big) \right.   \notag\\ 
    &\quad \left. + m_s \left(8 \zeta^2 m_s p_1^2 \left(42 d_1^3-576 b_4 (d_1 +6) +812 d_1^2 + 96 d_1 d_2  +288 \gamma_1^4 (d_1 +6)+ 600 d_2 \right.\right.\right.   \notag\\
    &\quad \left.\left.\left.  +4515 d_1 +7064\right)+3 \zeta^4 m_s^3 \left(3 \beta \left(4 d_1^3 +81 d_1^2 +6 d_1 d_2 +486 d_1  + 36 d_2 +872\right) \right.\right.\right.   \notag\\
    &\quad \left.\left.\left. -8 \lambda (8 d_1 +45)\right) +288 \sigma_s (d_1 +4)\right) +6 \zeta^4 m_u^4 \left(6 \beta \left(4 d_1^3 + 81 d_1^2 + 6 d_1 d_2 +36 d_2 \right.\right.\right.   \notag\\
    &\quad \left.\left.\left. + 486 d_1 +872\right) +\gamma^2 (11 d_1 +61) -8 \lambda (8 d_1 +45)\right) +576 m_u \sigma_u (d_1 +4)\right),   
    \\
    p_{4l} &= \frac{1}{2304} \beta  \zeta^2 (d_1 +4) \left(12\zeta^2 m_u^4 \left(3 \left(\gamma^2 -8  \lambda\right) +\beta \left(9 d_1^2  +108 d_1 +308\right)\right) \right.   \notag\\
    &\quad \left. +4 m_u^2 \big(3 \zeta^2 m_s^2 \left(6 \gamma^2 +\beta  \left(9 d_1^2 +108 d_1  +308\right)\right) +8 p_1^2 \left(9 d_1^2 +111 d_1 +9 d_2 +286\right) \right.   \notag\\
    &\quad \left. -9\sqrt{2} \gamma \zeta m_s p_1 (9 d_1 +52)\big) + 3\zeta^2 m_s^4 \left(\beta \left(9 d_1^2 + 108 d_1 +308\right) -48  \lambda \right) \right.   \notag\\
    &\quad \left. +16 m_s^2 p_1^2 \left(9 d_1^2 +111 d_1 +9 d_2  +286\right)\right).
\end{align}
Coefficients for $\chi_u(z)$:
\begin{align}
    c_{3l} &= -\frac{1}{288} \zeta  m_u \big(3\zeta^2 \left(2 m_u^2 \left(3 \left(\gamma^2 - 8  \lambda\right) +\beta \left(9 d_1^2  +108 d_1  +308 \right)\right) \right.   \notag\\
    &\quad \left. +m_s^2 \left(6 \gamma^2+\beta  \left(9 d_1^2 +108 d_1 +308\right)\right)\right) +16 p_1^2 \left(9 d_1^2 +111 d_1 + 9 d_2 +286\right)   \notag\\
    &\quad -12 \sqrt{2} \gamma  \zeta  m_s p_1 (9 d_1 +52)\big),   
    \\
    c_{4} &= \frac{1}{5184 \zeta} \left(\zeta m_u \left(-32 \zeta p_1^3 \left(-96 b_4 (4 d_1 +25) +36 d_1^3  +741 d_1^2 +90 d_1 d_2  \right.\right.\right.   \notag\\
    &\quad \left.\left.\left. +48 \gamma_1^4 (4 d_1 +25) + 4359 d_1 +567 d_2 +7342\right) +9 \sqrt{2} \gamma  \left(\zeta^4 m_s^3 \left(4 \left(\gamma^2 - 8 \lambda\right) \right.\right.\right.\right.   \notag\\
    &\quad \left.\left.\left.\left. + 3 \beta \left(7 d_1^2 + 94 d_1  +296\right)\right) +48 \sigma_s\right) -6 \zeta^3 m_s^2 p_1 \left(3\beta  \left(54 d_1^3 + 60 d_1 d_2  \right.\right.\right.\right.   \notag\\
    &\quad \left.\left.\left.\left. +1055 d_1^2  +6258 d_1  + 366 d_2 +11308\right) +2 \gamma^2 (66 d_1  +397)\right) +72 \sqrt{2} \gamma  \zeta ^2 m_s p_1^2 \right.\right.   \notag\\
    &\quad \left.\left.\left(26 d_1^2  +327 d_1 +14 d_2 +930\right) -1728 \zeta p_3 (d_1 +7)\right) -3 \zeta^4 m_u^3 \big(4 p_1 \left(3 \beta  \left(54 d_1^3 \right.\right. \right.   \notag\\
    &\quad +60 d_1 d_2   \left.\left.\left. +1055 d_1^2  +6258 d_1 +366 d_2 +11308\right) +\gamma^2 (66 d_1 +397)\right) \right.   \notag\\
    &\quad -24\lambda (22 d_1 +123) -3 \sqrt{2} \gamma \zeta m_s \left(2\left(9\gamma^2 -88  \lambda\right) +\beta \left(78 d_1^2  +997 d_1 +2956\big) \right)\right)   \notag\\
    &\quad + 432\sigma_u \big(\sqrt{2} \gamma \zeta  m_s -4 p_1 (d_1 +7)\big)\big),
    \\
    c_{4l} &= -\frac{1}{1728} \zeta  m_u \big(-32 p_1^3 (d_1 +7) \left(-96 b_4 +48 \gamma_1^4 + 9 d_2 +9 d_1^2  +111 d_1 +278\right)   \notag\\
    &\quad +3 \sqrt{2} \zeta ^3 m_s \left(\gamma  m_u^2 \left(9 \gamma ^2+\beta  \left(45 d_1^2  +561 d_1  +1708\right) -24  \lambda\right) \right.   \notag\\
    &\quad \left. +\gamma m_s^2 \left(3 \left(\gamma^2 - 8  \lambda\right) +\beta \left(9 d_1^2  +108 d_1 +308\right)\right)\right) - 6 \zeta^2 p_1 \left(2 m_u^2 \right.   \notag\\
    &\quad \left.\left(\beta  \left(27 d_1^3 +519 d_1^2 +18 d_1 d_2 +3190 d_1 +126 d_2 +6160\right) +\gamma^2 (12 d_1 +73)\right.\right.   \notag\\
    &\quad \left.\left. - 24  \lambda (d_1 +7)\right) + m_s^2 \left(\beta \left(27 d_1^3 +519 d_1^2 +18 d_1 d_2 +3190 d_1 +126 d_2 +6160\right)\right.\right.   \notag\\
    &\quad \left.\left. +2 \gamma^2 (12 d_1 +73)\right)\right) +8 \sqrt{2} \gamma \zeta  m_s p_1^2 \left(45 d^2 +567 d_1 +18 d_2  +1664\right)\big).
\end{align}
Coefficients for $\chi_s(z)$:
\begin{align}
    \tilde{c}_{3l} &= -\frac{1}{288} \zeta \big(2 m_s \left(3 \zeta^2 m_u^2 \left(6 \gamma^2 +\beta \left(9 d_1^2 +108 d_1 +308\right)\right) +8 p_1^2 \right.   \notag\\
    &\quad \left.\left(9 d_1^2 + 9 d_2 +111 d_1 +286\right)\right) +3 \zeta^2 m_s^3 \left(\beta \left(9 d_1^2 +108 d_1  +308\right)\right.   \notag\\
    &\quad \left. -48 \lambda\right) -12 \sqrt{2} \gamma  \zeta m_u^2 p_1 (9 d_1 +52)\big),
    \\
    \tilde{c}_{4} &= \frac{1}{5184 \zeta} \left(-4 \zeta ^2 m_s \left(8 p_1^3 \left(-96 b_4 (4 d_1 +25) +36 d_1^3 +741 d_1^2 +90 d_1 d_2 \right.\right.\right.   \notag\\
    &\quad \left.\left.\left.+48 \gamma_1^4 (4 d_1  +25)+4359 d_1 +567 d_2  +7342\right)+3 \zeta ^2 m_u^2 p_1 \left(3 \beta  \left(54 d_1^3  \right.\right.\right.\right.   \notag\\
    &\quad \left.\left.\left.\left. +1055 d_1^2 +60 d_1 d_2 +6258 d_1 +366 d_2 +11308\right)+2 \gamma ^2 (66 d_1  +397)\right)\right.\right.   \notag\\
    &\quad \left.\left.+432 p_3 (d_1  +7)\right)+9 \sqrt{2} \gamma  \zeta ^5 m_u^2 m_s^2 \left(2 \left(7 \gamma ^2-72  \lambda\right)+\beta  \left(57 d_1^2 +715 d_1 \right.\right.\right.   \notag\\
    &\quad \left.\left.\left. +2068\right)\right) +18\big(\sqrt{2} \gamma \zeta^5 m_u^4 \left(4 \left(\gamma^2 - 8\lambda\right) +3 \beta  \left(7 d_1^2 +94 d_1  +296\right)\right)\right.   \notag\\
    &\quad \left. + 4\sqrt{2} \gamma  \zeta^3 m_u^2 p_1^2 \left(26 d_1^2 +327 d_1+14 d_2  +930\right)-96 p_1 \sigma_s (d_1 +7)\right.   \notag\\
    &\quad \left. +48 \sqrt{2} \gamma \zeta m_u \sigma_u\big) -18 \zeta^4 m_s^3 p_1 \left(\beta \left(54 d_1^3  +1055 d_1^2 +60 d_1 d_2 +6258 d_1 \right.\right.\right.   \notag\\
    &\quad \left.\left.\left. +366 d_2 +11308\right) -16 \lambda (22 d_1  +123)\right)\right),   
    \\
    \tilde{c}_{4l} &= -\frac{1}{1728} \zeta \left(4 m_s \left(-8 p_1^3 (d_1 +7) \left(-96 b_4 +48\gamma_1^4 +9 d_1^2  +111 d_1 +9 d_2 \right.\right.\right.   \notag\\
    &\quad \left.\left.\left. +278\right) -3 \zeta^2 m_u^2 p_1 \left(\beta \left(27 d_1^3  +519 d_1^2 +18 d_1 d_2 +3190 d_1  +126 d_2 \right.\right.\right.\right.   \notag\\
    &\quad \left.\left.\left.\left. +6160\right) +2 \gamma ^2 (12 d_1  +73)\right)\right) +3 \sqrt{2} \zeta ^3 m_u^2 m_s^2 \left(6 \gamma^3 +\beta \gamma \left(36 d_1^2  +453 d_1  \right.\right.\right.   \notag\\
    &\quad \left.\left.\left. +1400\right)\right) +2 \sqrt{2} \zeta  m_u^2 \left(3 \gamma \zeta ^2 m_u^2 \left(3 \left(\gamma^2 -8  \lambda\right) +\beta \left(9 d_1^2  +108 d_1 \right.\right. \right.\right.   \notag\\
    &\quad \left.\left. \left.\left. +308\right)\right) +4 \gamma p_1^2 \left(45 d_1^2 +567 d_1 +18 d_2 +1664\right)\right)-6 \zeta ^2 m_s^3 p_1 (d_1 +7) \right.   \notag\\
    &\quad \left.\left(\beta \left(27 d_1^2 +330 d_1 +18 d_2 +880\right) -48 \lambda\right)\right).
\end{align}

\section{Chiral Condensates via Holographic Renormalization} \label{app3}

We compute the chiral condensates in the 2+1-flavor EDF system using holographic renormalization. The procedure begins by constructing the on-shell action, incorporating the Gibbons-Hawking term to guarantee a well-defined Dirichlet variational problem and boundary counterterms that cancel UV divergences, following the standard method outlined in \cite{Skenderis:2002wp,deHaro:2000vlm}. The boundary action is given by
\begin{align}
   S_\partial &= \frac{1}{2\kappa_N^2} \int_{z\to 0} dx^4\sqrt{-h} \left[2K -\frac{4}{3} \phi^2 -6 -\beta \left(\frac{1}{2}\left(\chi_u^2+\chi_d^2 +\chi_s^2\right) -\frac{\gamma\chi_u\chi_d\chi_s}{2\sqrt{2}}\right. \right.   \notag\\
    &\quad \left.\left.  +(3+\frac{d_1}{2})\left(\phi\chi_u^2+\phi\chi_d^2+\phi\chi_s^2\right)\right) -\frac{16}{27} (1 +12b_4 -6\gamma_1^4) \phi^4\ln z +\beta \left(\frac{1}{18}(286 \right. \right.   \notag\\
    &\quad \left.\left. +111 d_1 + 9 d_2 +9 d_1^2)(\phi^2\chi_u^2+\phi^2\chi_d^2 +\phi^2\chi_s^2) -\frac{(52 +9d_1)\gamma}{6\sqrt{2}} \phi\chi_u\chi_d\chi_s \right.\right.   \notag\\
    &\quad \left.\left. +\frac{1}{96}(6\gamma^2+\beta(308 +108d_1 + 9 d_1^2 )) (\chi_u^2 \chi_d^2+\chi_u^2 \chi_s^2+\chi_d^2 \chi_s^2) + \frac{1}{192}(\beta(308\right.\right.   \notag\\
    &\quad \left.\left. +108d_1 +9 d_1^2 ) -48\lambda) (\chi_u^4+\chi_d^4+\chi_s^4)\right) \ln z - k \phi^4\right],
\end{align}
where $h$ is the determinant of the induced boundary metric, and $K$ is the trace of the extrinsic curvature. The constant $k$ is introduced to enforce $p(T=0)=0$. Note that we have $\chi_u =\chi_d$ for the $2+1$--flavor case. Varying the renormalized on-shell action with respect to the quark masses yields the chiral condensates:
\begin{align}
    \langle\Bar{\psi} \psi\rangle_{u}^T &= \frac{\delta (S +S_\partial)_{on-shell}}{2\delta m_u}   \notag\\
    &= \frac{1}{2\kappa_N^2} \lim_{z\to 0} \sqrt{-h}\,z\,\zeta \left(-\beta e^{\phi} n_\mu\nabla^\mu\chi_u-\beta\chi_u -\beta \left(6 +d_1\right) \phi\chi_u +\frac{\beta\gamma}{2\sqrt{2}} \chi_u\chi_s\right.   \notag\\
    &\quad \left. -\frac{\beta\gamma \left(52 +9d_1\right)}{6\sqrt{2}} \phi\,\chi_u\,\chi_s\ln z +\frac{\beta}{9}  \left(286 +111d_1 +9d_2 +9 d_1^2\right) \phi^2\chi_u\ln z \right.   \notag\\
     &\quad \left. +\frac{\beta}{48} \left(6\gamma^2 +\beta \left(308 +108d_1 +9d_1^2\right)\right) \chi_u\,\chi_s^2 \ln z \right.   \notag\\
   &\quad \left. +\frac{\beta}{24} \left(\beta\left(308 +108 d_1 +9 d_1^2\right) +3\left(\gamma^2 -8\lambda \right)\right) \chi_u^3\ln z\right),\label{psibarpsiuR}
     \\
\langle\Bar{\psi} \psi\rangle_{s}^T &=\frac{\delta (S +S_\partial)_{on-shell}}{\delta m_s}   \notag\\
     &= \frac{1}{2\kappa_N^2} \lim_{z\to 0} \sqrt{-h}\,z\,\zeta \left(-\beta e^\phi n_\mu \nabla^\mu\chi_s -\beta\chi_s -\beta \left(6 +d_1\right) \phi\chi_s +\frac{\beta\gamma}{2\sqrt{2}} \chi_u^2\right.   \notag\\
     &\quad \left. -\frac{\beta\gamma (52+9d_1)}{6\sqrt{2}} \phi\,\chi_u^2\ln z +\frac{\beta}{9} \left(286 +111d_1 +9d_2 +9 d_1^2\right)\phi^2 \chi_s \ln z \right.   \notag\\
     &\quad \left. +\frac{\beta}{24} \left(6\gamma^2 +\beta\left(308 +108 d_1 +9d_1^2\right)\right)\chi_u^2\,\chi_s\ln z \right.   \notag\\ 
    &\quad \left. +\frac{\beta}{48} \left(\beta \left(308 +108 d_1 +9 d_1^2\right) -48\lambda\right) \chi_s^3 \ln z\right),
\end{align}
where $n^\mu$ denotes the outward-pointing unit normal vector at the boundary. Substituting the UV expansions \eqref{UV_f}--\eqref{UV_chi_s} gives
\begin{align}
  \langle\Bar{\psi} \psi\rangle_{u}^T &= \frac{\beta}{\kappa_N^2} \sigma_{u} +b_1   \notag\\
    &= \frac{\beta}{\kappa_N^2} \sigma_u -\frac{1}{1728 \kappa_N^2} \big(\beta \zeta^2 m_u \big(3\zeta^2 \left(2 m_u^2 \left(9 \left(3 \gamma^2-8 \lambda\right) +\beta  \left(81 d_1^2 +918 d_1 +2536\right)\right)\right.   \notag\\
    &\quad \left. +m_s^2 \left(54 \gamma^2 +\beta \left(81 d_1^2 +918 d_1 +2536\right) \right)\right) +16 p_1^2 \left(81 d_1^2 +819 d_1 + 27 d_2  +1907\right)   \notag\\
   &\quad -36 \sqrt{2} \gamma \zeta  m_s p_1 (27 d_1  +136)\big)\big) ,
\\
\langle\Bar{\psi}\psi\rangle_{s}^T &= \frac{\beta}{\kappa_N^2} \sigma_{s} +b_2   \notag\\
    &= \frac{\beta}{\kappa_N^2} \sigma_s -\frac{1}{1728 \kappa_N^2} \big(\beta \zeta^2 \big(2 m_s \left(3 \zeta^2 m_u^2 \left(54 \gamma^2 +\beta \left(81 d_1^2 +918 d_1 +2536\right)\right) \right.   \notag\\
    &\quad \left. +8 p_1^2 \left(81 d_1^2 +819 d_1 + 27 d_2 +1907\right)\right) +3 \zeta^2 m_s^3 \left(\beta \left(81 d_1^2 +918 d_1  +2536\right)\right.   \notag\\
    &\quad \left. -144 \lambda\right) -36 \sqrt{2} \gamma \zeta m_u^2 p_1 (27 d_1 +136)\big)\big).
\end{align}

\bibliography{ref}

\end{document}